# The Unusual Apparition of Comet 252P/2000 G1 (LINEAR) and Comparison with Comet P/2016 BA$_{14}$ (PanSTARRS)

Jian-Yang Li (李荐扬)[1], Michael S.P. Kelley[2], Nalin H. Samarasinha[1], Davide Farnocchia[3], Max J. Mutchler[4], Yanqiong Ren (任彦琼)[5], Xiaoping Lu (盧曉平)[5], David J. Tholen[6], Tim Lister[7], Marco Micheli[8]

[1] Planetary Science Institute, 1700 E. Ft. Lowell Rd., Suite 106, Tucson, AZ 85719, USA; jyli@psi.edu, nalin@psi.edu
[2] Department of Astronomy, University of Maryland, College Park, MD 20742, USA; msk@astro.umd.edu
[3] Jet Propulsion Laboratory, California Institute of Technology, Pasadena, CA 91109, USA; davide.farnocchia@jpl.nasa.gov
[4] Space Telescope Science Institute, 3700 San Martin Drive, Baltimore, MD 21218-2463, USA; mutchler@stsci.edu
[5] Macau University of Science and Technology, Macau, China; xplu@must.edu.mo, yanqiongren@qq.com
[6] Institute for Astronomy, University of Hawai'i, Honolulu, HI 96822, USA; tholen@IfA.Hawaii.Edu
[7] Las Cumbres Observatory, 6740 Cortona Drive Ste. 102, Goleta, CA 93117, USA; tlister@lco.global
[8] ESA SSA-NEO Coordination Centre, 00044 Frascati (RM), Italy; marco.bs.it@gmail.com

**Abstract:**

We imaged Comet 252P/2000 G1 (LINEAR) (hereafter 252P) with the *Hubble Space Telescope* and both 252P and P/2016 $BA_{14}$ (PanSTARRS) (hereafter $BA_{14}$) with the *Discovery Channel Telescope* in March and April 2016, surrounding its close encounter to Earth. The *r'*-band *Afρ* of 252P in a 0.2"-radius aperture were 16.8±0.3 and 57±1 cm on March 14 and April 4, respectively, and its gas production rates were: $Q(OH) = (5.8±0.1)\times10^{27}$ $s^{-1}$, and $Q(CN) = (1.25±0.01)\times10^{25}$ $s^{-1}$ on April 17. The *r'*-band upper limit *Afρ* of $BA_{14}$ was 0.19±0.01 cm in a 19.2"-radius aperture, and $Q(CN) = (1.4±0.1)\times10^{22}$ $s^{-1}$ on April 17, 2017. 252P shows a bright and narrow jet of a few hundred kilometers long in the sunward direction, changing its projected position angle in the sky with a periodicity consistent with 7.24 hours. However, its photometric lightcurve is consistent with a periodicity of 5.41 hours. We suggest that the nucleus of 252P is likely in a non-principal axis rotation. The nucleus radius of 252P is estimated to be about 0.3±0.03 km, indicating an active fraction of 40% to >100% in its 2016 apparition. Evidence implies a possible cloud of slow-moving grains surrounding the nucleus. The activity level of 252P in the 2016 apparition increased by two orders of magnitude from its previous apparitions, making this apparition unusual. On the other hand, the activity level of $BA_{14}$ appears to be at least three orders of magnitude lower than that of 252P, despite its ten times or larger surface area.

*Keywords:* comets: individual (252P/2000 G1 (LINEAR), P/2016 $BA_{14}$ (PanSTARRS)) – methods: observation – techniques: high angular resolution – techniques: photometric

## 1. Introduction

Comet 252P/2000 G1 (LINEAR) (hereafter 252P) is special among Jupiter family comets in two aspects. First, it is probably among the smallest known Jupiter family comets (JFCs) as suggested by previous observations. The observations from the discovery apparition (Shelly et al. 2000) showed a very low dust production rate with an *Afρ* of ~2 cm (*Afρ* is an empirical estimate of cometary activity, see A'Hearn et al. 1984) near its perihelion, which is at least one order of magnitude lower than typical Jupiter family comets. Assuming that the dust production rate scales with surface area, even for a small active area of 0.2% (such as 10P/Tempel 2, A'Hearn et al. 1995), the nucleus radius has to be <0.9 km. This size is consistent with its brightness at large heliocentric distances when the cometary activity is absent or weak, putting 252P in the small end of the size distribution of known JFC nuclei (c.f., Fernández et al. 2013).

Since comets are the original building blocks of outer solar system bodies, if the size distribution of pristine cometary nuclei is primordial, then it is related to the fundamental properties of cometesimals that are determined by the physical conditions and accretion processes during the formation of planetary systems. However, the surfaces and shapes of JFCs may have highly evolved by mass loss during their numerous perihelion passages, mostly within 1.5 au from the Sun. The evolution of JFCs is not well understood. The observations by the Deep Impact and Stardust-NExT flybys for Comet 9P/Tempel 1 (hereafter 9P) (Thomas et al. 2007, 2013) and by the Rosetta for Comet 67P/Churyumov-Gerasimenko (hereafter 67P) (Groussin et al. 2015) revealed some erosional processes in localized areas on the nuclei. In general, comets smaller than ~1 km in radius are likely severely eroded based on the assumption that a single power law should fit the JFC cumulative size distribution (Meech et al. 2004, Lamy et al. 2004, Snodgrass et al. 2011). If the size of 252P is indeed small, then it is important to determine whether 252P is a highly evolved comet near the end of its life, or its nucleus is born small and representative of the small size end of the primordial cometesimal size distribution.

The second special aspect of Comet 252P is that it could be a split pair with Comet P/2016 BA$_{14}$ (PanSTARRS) (hereafter BA$_{14}$). BA$_{14}$ was initially thought to be an asteroid as the first observations upon discovery did not show cometary activity. Later observations revealed an extremely weak coma, and so BA$_{14}$ was reclassified as a comet (Knight et al. 2016). There is a remarkable similarity between the orbits of 252P and BA$_{14}$. Since the two comets both experience close encounters to Jupiter, the orbital elements significantly change over time. The classical orbital distance criteria (e.g., that of Southworth & Hawkins 1963) are not conclusive on whether or not these two objects are dynamically related. Of special interest is the Jupiter Tisserand parameter, which remains quasi-constant even with intervening planetary encounters. The values for 252P and that of BA$_{14}$ are 2.82 and within 1% of each other. Linking these two objects together in a detailed dynamical study is complicated by non-gravitational forces and close encounters with planets. But if confirmed, then the comparisons between these two comets could help understand the fragmentation process and evolution of cometary nuclei.

Comet 252P made a close encounter with Earth at 0.0356 au, or ~14 lunar distances on March 21.55211, 2016, the 5$^{th}$ closest approach of a comet to Earth ever known[1]. The small geocentric distance coupled with the high angular resolution of the *Hubble Space Telescope* (*HST*) *Wide Field Camera 3* (*WFC3*) provided us with a rare opportunity to observe 252P at a pixel scale as small as 1.5 km pixel$^{-1}$, facilitating the study of the fine structure in its inner coma and to infer the properties of its nucleus.

---
[1] https://cneos.jpl.nasa.gov/ca/, accessed on May 26, 2017

## 2. Observations

We observed 252P at two separate epochs on March 14 and April 4, 2016 (hereafter March epoch and April epoch, respectively) with the *HST/WFC3 UVIS* channel through the F555W (*V*-band) and F625W (*r'*-band) filters using mostly a 1k×1k subarray or 40"×40" field of view (HST program ID 14103). The March 14 observations consist of one *HST* orbit and 18 images, covering a total duration of 43 min. The April 4 observations have five *HST* orbits, 14 images per orbit, with the first four orbits executed continuously, and the last orbit delayed by about 1.5 hours from the fourth orbit. The total duration covered by the April epoch is 8.5 hours. Table 1 summarizes our observations and some basic measurements.

On April 17, 2016, we imaged both 252P and BA$_{14}$ with the *Large Monolithic Imager* (*LMI*) on Lowell Observatory's *Discovery Channel Telescope* (*DCT*), a 4.3 m aperture telescope located near Happy Jack, AZ. The 6k×6k pixel *LMI* chip was operated in 2x2 binning mode (0.24" per binned pixel), providing a full field of view of 12'×12'. The filter wheel was equipped with *V* and *r'* filters, as well as the *OH* (310 nm central wavelength), *CN* (387 nm), *BC* (445 nm), and *RC* (713 nm) comet narrowband filters from the HB filter set (Farnham et al. 2000). All four narrowband filters have a 6 nm FHWM. The HB filters are smaller than the standard *LMI* filters and vignette the field of view, limiting the usable area to an approximately 12' diameter circle. The weather was clear, but the seeing was poor, with a stellar FWHM of 2.2". Table 2 summarizes our *DCT* observations.

**Table 1**. Summary of observations and photometric measurements of 252P from *HST*.

| Epoch | March 14 | | April 4 | |
|---|---|---|---|---|
| Start UT | 2016-03-14T05:47 | | 2016-04-04T16:08 | |
| End UT | 2016-03-14T06:32 | | 2016-04-05T00:48 | |
| Heliocentric Distance (au) | 0.996 | | 1.037 – 1.038 | |
| Geocentric Distance (au) | 0.0568 – 0.0567 | | 0.0922 – 0.0942 | |
| Phase angle (degree) | 86.5 | | 64.0 – 64.5 | |
| Pixel Scale at Comet (km) | 1.65 | | 2.67 – 2.73 | |
| Filter | F555W | F625W | F555W | F625W |
| Effective Wavelength (nm) | 530.8 | 624.1 | 530.8 | 624.1 |
| Equivalent Width (nm) | 51.7 | 45.1 | 51.7 | 45.1 |
| Magnitude (VEGA=0) | 16.68 | 16.20 | 15.95 | 15.51 |
| $A(\alpha)f\rho$ (cm) | 15.5±0.3 | 16.8±0.3 | 54±1 | 57±1 |
| $A(0)f\rho$ (cm) | 29±0.6 | 30±0.6 | 151±3 | 160±3 |

Notes:
1. 2% absolute photometric error is estimated for $Af\rho$ and magnitude
2. Magnitude and $Af\rho$ are all for 0.2"-radius (5 pixels) circular aperture

**Table 2**. *Discovery Channel Telescope* observing circumstances on April 17, 2016.

| Target | Filters | Time (UT) | Airmass | $r_h$ [1] (au) | $\Delta$ [2] (au) | $\alpha$ [3] (°) |
|---|---|---|---|---|---|---|
| 252P/2000 G1 | CN, RC, BC | 08:35 | 1.46 | 1.098 | 0.164 | 51.3 |
|  | V, r' | 09:10 | 1.32 | 1.098 | 0.164 | 51.3 |
|  | r', OH, CN, RC, BC | 11:30 | 1.15 | 1.099 | 0.164 | 51.2 |
| P/2016 BA$_{14}$ | r', CN | 10:10 | 1.01 | 1.106 | 0.207 | 55.6 |

Notes:
[1] Heliocentric distance
[2] Geocentric distance
[3] Solar phase angle

## 2.1 Ephemeris update for HST observations

The planned observations of the comet near its close encounter at <0.1 au, coupled with the small field of view (FOV) of *WFC3/UVIS* detector (160"×160" maximum, as small as 40"×40" for our subarray images) posed a challenge for the accurate ephemeris predictions, especially for a comet. Reliable characterization of the comet's ephemeris prediction uncertainties was essential for the success of our HST observations.

Previous experience (Rayman 2003, Chesley & Yeomans 2006, Farnocchia et al. 2016) has shown the challenges of accurately estimating comet trajectories. The presence of a coma makes it difficult to correctly locate a comet's nucleus in optical images, and tailward (or sunward in some cases, e.g., Yeomans & Chodas 1989) biases can affect the quality of the astrometric positions used to estimate a comet's trajectory (Tholen & Chesley 2004). Moreover, non-gravitational accelerations could affect the trajectory in ways that are difficult to separate from astrometric biases. Generally, as the prediction lead times shorten and data arcs lengthen, the effects of modeling errors are significantly reduced. A further complication arises from the need for sufficient time to build and upload the *HST* observation sequence. This requirement places an ephemeris deadline roughly two weeks before the observations are executed. In two weeks, a comet could behave in unexpected ways, especially if close to perihelion, and we have no way to correct the trajectory accordingly.

The data arc for comet 252P is composed of three apparitions, 2000, 2011, and 2015-2016. To mitigate the effect of astrometric biases we conducted an observation campaign from Mauna Kea, *Las Cumbres Observatory*, and the *DCT* to obtain high-quality imaging data. This observational effort includes the September 2015 recovery of the comet (MPEC 2015-S97). The astrometric positions were extrapolated to the limit of a zero-sized photometric aperture around the brightness peak (Tholen & Chesley 2004), and we also estimated their uncertainties when estimating the trajectory of 252P. For the 2015 – 2016 interval, we only used astrometry from our observation campaign.

To account for non-gravitational perturbations, we considered two models. The first one is the classical Marsden et al. (1973) model, where the radial, transverse, and normal acceleration components are $A_i g(r)$, $i$=1, 2, 3, and $g(r)$ is a scale factor driven by water sublimation with a peak at perihelion. An asymmetric variant of this model includes a time-offset $\Delta T$ in the non-gravitational acceleration peak with respect to perihelion (Yeomans & Chodas 1989). A second, higher fidelity model is the rotating jet model (Chesley & Yeomans 2005), which averages the thrust of a discrete number of jets over a comet rotation.

To analyze the ephemeris of 252P we computed solutions corresponding to three configurations:
1) Data since 2000, Marsden et al. (1973) model with $A_1$, $A_2$, $A_3$ and $\Delta T$.
2) Data since 2000, rotating jet model with two jets.
3) Data since 2011, Marsden et al. (1973) model with $A_1$, $A_2$, $A_3$.

These sets of solutions were helpful to compare the different non-gravitational models and to analyze the *sensitivity* to the length of the data arc. In particular, while a long data arc put a better formal constraint on the trajectory, non-gravitational perturbations could change behavior with time and so the corresponding ephemeris could be negatively affected.

Figure 1 shows the comparison between the different predictions for the geocentric ephemeris at closest approach. The data cutoff for this prediction is March 7, two weeks before perigee. The jet model and the short-arc solutions provide successful predictions as the reconstructed astrometric position (computed using post-encounter data) falls well within the uncertainty ellipses. Between these two solutions the jet model has smaller prediction uncertainties thanks to the longer data arc used in the fit. On the other hand the $A_1$, $A_2$, $A_3$ and $\Delta T$ solution over the long arc fails to accurately predict the plane-of-sky position of the comet at closest approach. This result further confirms the suitability of the rotating jet model and that it is important to exercise care when comet trajectories are fit to a long data arc.

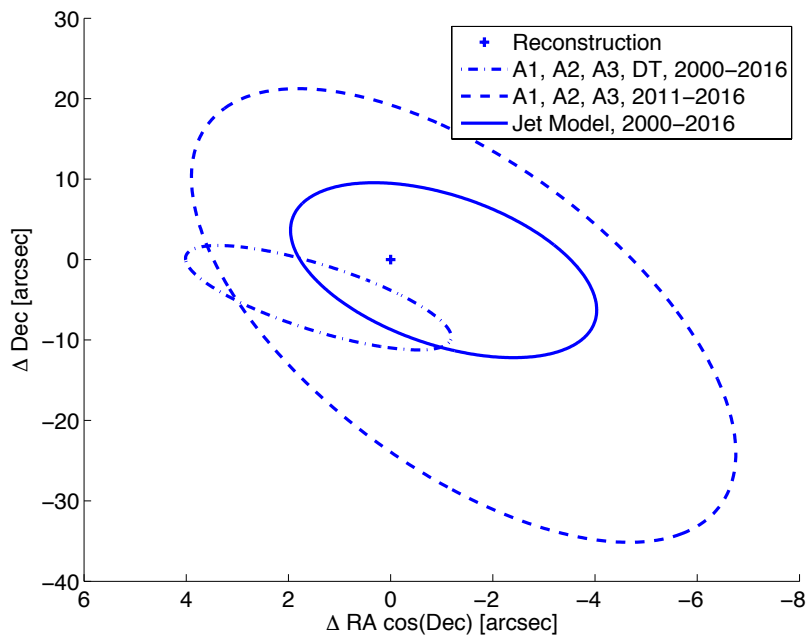

**Figure 1**: Geocentric plane-of-sky predictions at closest approach on 2016-03-21 at 13:15 TDB obtained with different models for non-gravitational perturbations and different data arcs. The origin is marked with a plus sign and corresponds to the reconstructed astrometric position of RA = 255.3216° and Dec = -72.9076°.

Another useful step to assess the consistency of an orbit solution is to track its evolution as more data become available to refine the orbit. Figure 2 shows this evolution for the jet model solution starting from a data cutoff of January 27, 2016 and updating the solution every ten days until March 7, 2016. The prediction ellipses are nicely nested and converge toward the final reconstruction, thus increasing our confidence that the prediction was robust. For these reasons we used the jet model solution for the *HST* pointing.

As will be discussed later in Sections 3.1 and 4.2, 252P experienced a significant increase in brightness and non-gravitational perturbations around perihelion. None of the solutions discussed above can fit the post-perihelion data and actually we have not yet found a way to fit the whole 2015-2016 apparition data. When planning comet observations, it is wise to have some margin to account for this stochastic behavior in non-gravitational perturbations, especially when the ephemeris delivery is a week or more before the observation execution. The experience with the ephemeris predictions of 252P for the *HST* observations is applicable to future comet observations using small-FOV detectors, especially near their close encounters with the observatory.

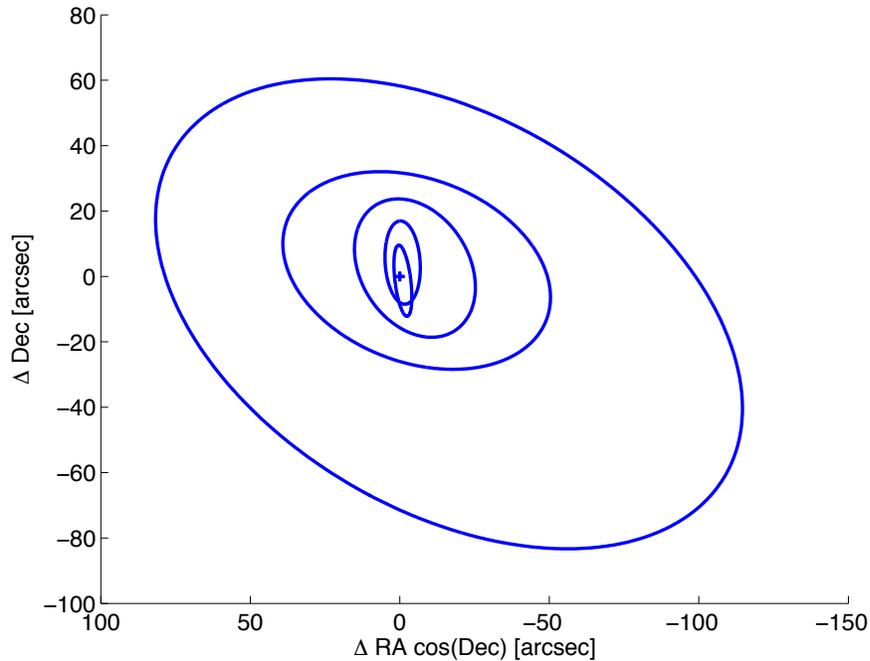

**Figure 2**: Evolution of the geocentric plane-of-sky prediction at closest approach on 2016-03-21 at 13:15 TDB obtained with the rotating jet model. The ellipses are 3-σ and shrink as the data cutoff advances from 2016-01-27 to 2016-03-07 with a time step of ten days. The origin is marked with a plus sign and corresponds to the reconstructed astrometric position of RA = 255.3216° and Dec = -72.9076°.

*2.2 Image reduction*
*2.2.1   HST/WFC3 data*

All images are reduced by the standard *HST/WFC3* calibration pipeline at Space Telescope Science Institute (STScI) (Deustua 2016). We estimated the sky background from boxes of 100×100 pixels in size at the four corners of each image that have their corners 20 pixels away from the image corners. Inspecting images and comparing the background levels measured from the four boxes suggested that, for the March epoch, the coma probably filled the whole 40" image frame except for the corner to the south of the comet (the Sun is to the west and tail to the east). We therefore only used the measurement from the south corner as the sky background estimate for images in this epoch. For the April epoch, on the other hand, the background is elevated in the box in the projected cometary tail (west) direction in all images, while similar in all other three boxes. We therefore used the average from the other three boxes, i.e., in the sunward corner and the two corners on the sides of the comet, as sky background. We then subtract the sky background from all images for future analysis.

The centroid is measured by fitting a 2D Gaussian to the center core of the comet in a 6×6 pixel box. We experimented with various sizes for the box and observed that the best-fit center shifted by <0.1 pixel in >90% images, and up to 0.5 pixel for a few. However, about 20 images show obvious linear smearing of up to 7 pixels (see Section 2.3). For those images, the centroid usually lands near the middle of the elongated core. Caution has to be exercised, though, in interpreting the results derived from smeared images using the corresponding centroid measurements.

We then measured the total counts within circular apertures centered at the centroid and with 1-300 pixels in radii, and converted to flux and magnitude in Vega-magnitude system following the photometric calibration constants[2] (Deustua et al. 2016). Table 1 lists the average magnitude of 252P at the two epochs through the two filters that we used.

*2.2.2   DCT/LMI*

We subtracted the bias and applied flat-field corrections created from twilight sky to all images. The photometric calibration was generated from observations of standard stars from the Landolt (2009), Smith et al. (2002), and Farnham et al. (2000) catalogs. Extinction across the *OH* filter bandpass is highly variable and critically depends on the ozone content of the atmosphere. Therefore, we followed the iterative procedure for *OH* calibration outlined in Farnham et al. (2000). Images were median combined in the rest frame of each target (Figures 3 and 4).

The target brightnesses were measured with standard aperture photometry techniques using circular apertures up to 80-pixels (19.2") in radius. The Farnham et al. (2000) and Landolt (2009) catalogs are on the Vega magnitude system, whereas the Smith et al. (2002) catalog uses the AB magnitude system. The HB filter photometry was converted to gas flux (*CN*, *OH*) or continuum flux density (*BC*, *RC*) using the calibration constants from Farnham et al. (2000). For *V*, we used $3.631 \times 10^{-8}$ W m$^{-2}$ $\mu$m$^{-1}$ for the zero-point flux density and -26.76 for the apparent magnitude of the Sun (Bessel 1998). For *r'* we used an apparent magnitude of -26.93 for the

---

[2] http://www.stsci.edu/hst/wfc3/analysis/uvis_zpts/, accessed on May 26, 2017

Sun, based on the color of the Sun (*V-R*=0.370; Colina et al. 1996) and the Smith et al. (2002) *V* to *r'* transformation.

We converted observed fluxes into production rates using D. Schleicher's online tool at Lowell Observatory[3]. The calculations use a Haser (1957) model with scale-lengths from A'Hearn et al. (1995) and Cochran & Schleicher (1993). For 252P, we remove the dust continuum from the *OH* and *CN* filters by scaling the *BC* photometry to each respective filter using the continuum color. We measured a *BC-RC* color of 0.082 and 0.073 mag per 0.1 $\mu$m (centered at *BC*) in 40- and 80-pix apertures, respectively. For $BA_{14}$, we do not have a continuum color estimate, but experimented with a neutral reflectance, and a reddened color of 0.070 mag per 0.1 $\mu$m (similar to the measured values for 252P). The difference between the two colors is 6%, therefore we incorporate an additional uncertainty of 3% in our *CN* fluxes. Gas production rates and coma *Af*ρ values are presented for both comets in Table 3. Note that no dust coma for $BA_{14}$ was identifiable in the *r'*-band image, so its *Afρ* estimate should be treated as an upper limit.

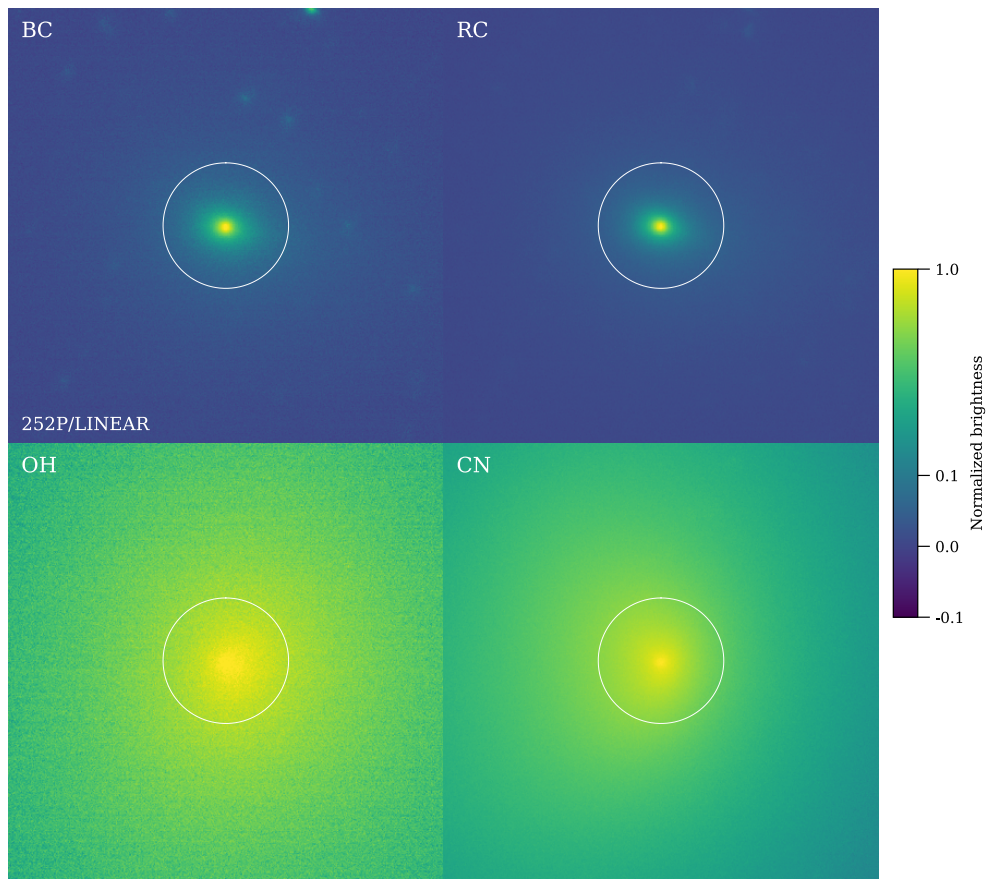

**Figure 3**: Comet 252P imaged by the *Discovery Channel Telescope* through the *BC*, *RC*, *OH*, and *CN* filters on 2016 April 17.48 UT. Each image covers 20,000×20,000 km at the distance of the comet, and an 80-pixel (19.2") circular aperture is shown. The images are normalized by the brightness of the center pixel, and displayed on a two-segment brightness stretch: a linear brightness stretch from -0.1 to 0.1, and a logarithmic stretch from 0.1 to 1.0. Celestial north is up, and east to the left.

---

[3] http://asteroid.lowell.edu/comet/cover_ghq.html, accessed on May 26, 2017

**Table 3**. *DCT/LMI* photometry and derived parameters.

| Target | Time (UT) | V m (mag) / Afρ (cm) [2] | r' m (mag) / Afρ (cm) [2] | OH m (mag) / Q ($10^{25}$/s) [1] | CN m (mag) / Q ($10^{23}$/s) [1] | BC m (mag) / Afρ (cm) [2] | RC m (mag) / Afρ (cm) [2] |
|---|---|---|---|---|---|---|---|
| 252P/2000 G1 | 08:35 | | | | 10.30 (0.01) / 120 (1) | 13.06 (0.01) / 24.3 (0.2) | 11.59 (0.01) / 28.9 (0.3) |
| 252P/2000 G1 | 09:10 | 12.01 (0.02) / 39.5 (0.7) | 11.97 (0.03) / 34.9 (1.0) | | | | |
| 252P/2000 G1 | 11:30 | | 11.99 (0.03) / 34.6 (1.0) | 9.88 (0.02) / 581 (11) | 10.26 (0.01) / 125 (1) | 13.04 (0.01) / 24.8 (0.2) | 11.57 (0.01) / 29.7 (0.3) |
| P/2016 $BA_{14}$ | 10:10 | | 17.90 (0.04) / 0.19 (0.01) | | 17.56 (0.10) / 0.14 (0.01) | | |

Notes:
All photometry is within a 19.2" radius circle. Uncertainties are empirically measured and listed in the parentheses. Calibration uncertainties for the HB filters and derived gas fluxes are 5 to 10% (Farhnam et al. 2000).
[1] Molecule production rate
[2] *Afρ* parameter of A'Hearn et al. (1984)

## 2.3 Smearing in HST/WFC3 images

Since 252P was <0.1 au from Earth during our observations, the orbital motion of *HST* resulted in a parallax of the target of up to 500", or >3 FOVs of *UVIS* full frame. While *HST* automatically compensated for the parallax in its pointing based on the calculated orbital position, the comet still drifted inside the FOV, causing smearing in many frames. To quantify the smearing, we used the time-rate of the measured centroid locations within each *HST* orbit to approximate the drift and direction of the pointing at the time of each image acquisition. For the exposure time, we can calculate the corresponding amount and the direction of smearing in each image frame (Figure 5). For about half of the images, the smearing is <2 pixels, while in about ten images the smearing is >5 pixels and up to 7 pixels.

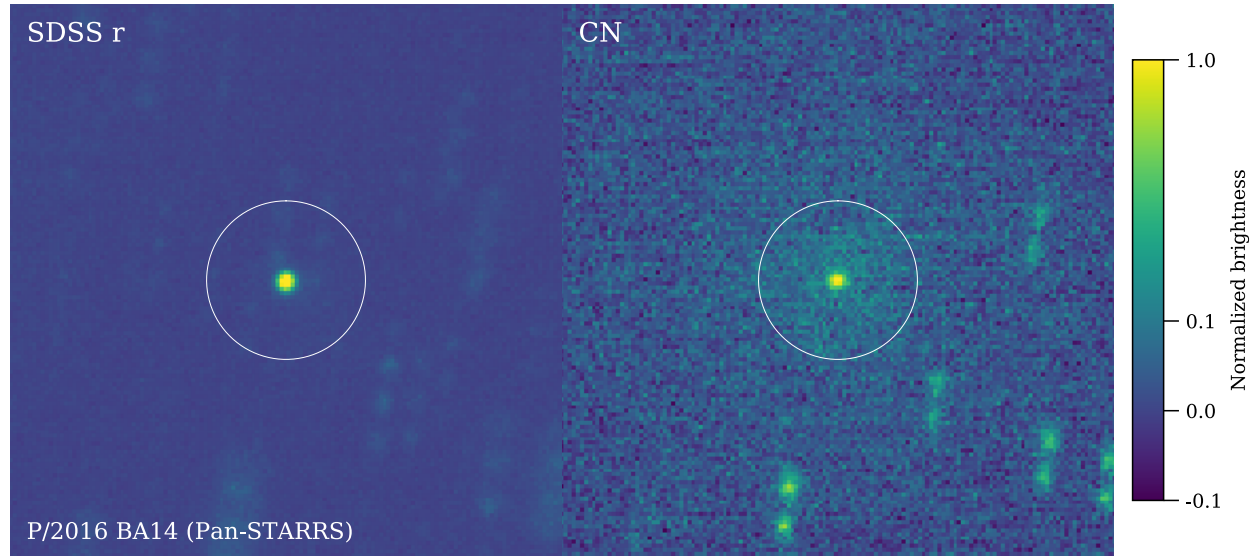

**Figure 4**: Same as Figure 3, but for $BA_{14}$ and only for the *r'* and *CN* filters. The images were rebinned by a factor of 4 before plotting to enhance the faint CN coma.

Smearing affects the measurement of the centroid and all quantities that rely on it, such as aperture photometry, color, and coma morphology. The various amounts of smearing in our images suggests that we should not interpret those measurements made from <5 pixel or smaller apertures or distances from the center without taking special precautions. In addition, we should check all results against the possible effects due to smearing in order not to interpret artifacts.

*2.4 Saturation*

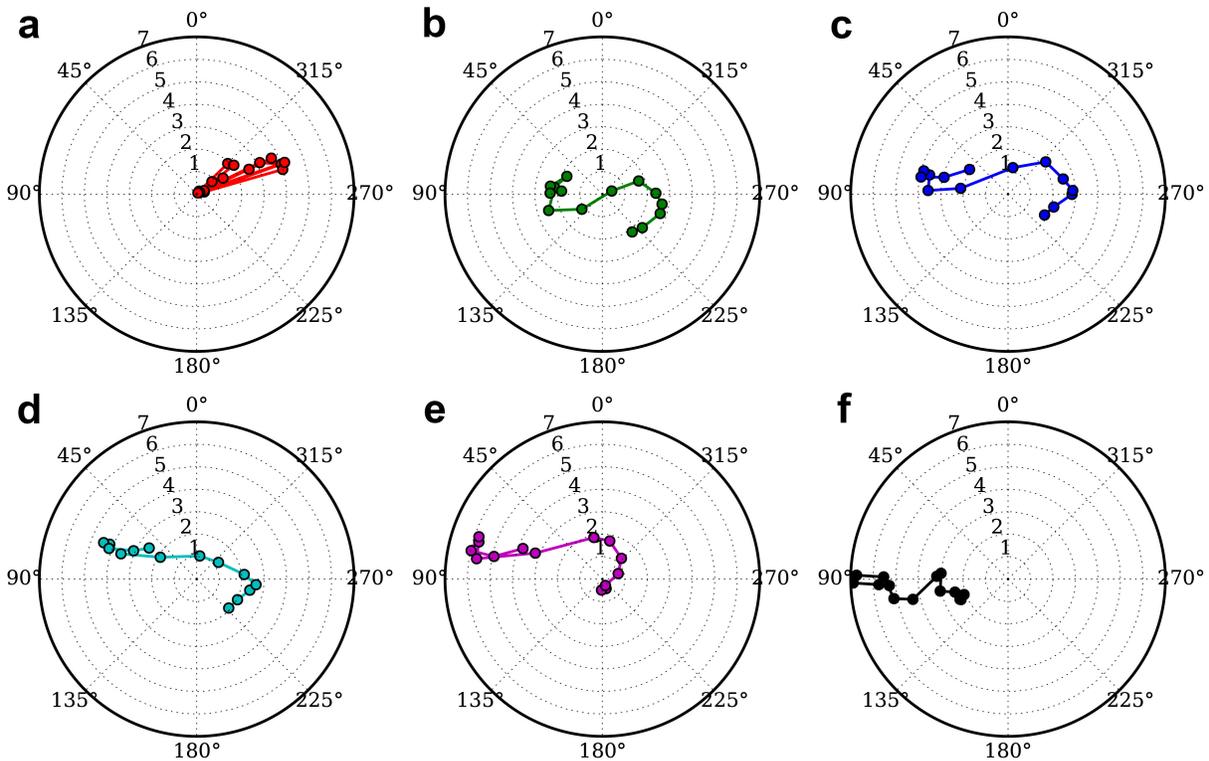

**Figure 5**: The smearing of all images. The polar coordinates of symbols mark the amplitudes (radial coordinate) and the directions (azimuthal coordinate) of the smearing in each image. The amplitude of smearing is measured in pixels (marked by numerals 1 through 7), and the direction of smearing is measured in degrees from the up direction in the original image frame, increasing counter-clockwise. The three upper panels from left to right correspond to the March 14 orbit and the first and second orbits in the April epoch, and the three lower panels from left to right correspond to the third, fourth, and fifth orbits in the April epoch. The small smearing of ~0 for some images in the March 14 orbit is due to short exposure time of 5 s. All images in the April epoch have 80 s exposure time. The images in the last orbit have the largest smearing, with a minimum of 2 pixels and a maximum of 7 pixels.

Five images have their peak pixel values above the 67k DN linearity limit for the detector we used (UVIS2 C-amplifier) with a maximum of 71k DN. Aperture photometry can still be preserved with great accuracy as long as all the blooming in the pixels adjacent to the saturated pixel is included in the measurement (Deutua 2016). The blooming is expected to be small for our maximum saturation level of 6% above linearity limit in a single pixel, and therefore no detectable effect in photometry is expected. For nucleus size measurements (Section 3.4), we simply avoided using those five images with saturated peak pixels.

## 3. Results
*3.1 Lightcurve*

The lightcurve of the comet within a 10-pixel (27 km) radius aperture on April 4 shows clear variability (Figure 6 upper) that appears to contain both a periodic component and an overall decreasing trend in a timescale much longer than the duration of our observations in this epoch. The photometric data of 252P from the Minor Planet Center (MPC) database suggest that the comet increased its total brightness significantly around the start of March 2016 with an extremely steep dependence on heliocentric distance. Given the small changes in the geometry of 252P during the April epoch (~0.1% in heliocentric distance, 2% in the geocentric distance, and 0.44º in phase angle), we decided to correct the photometry for its geometry dependence using a simple linear correction with respect to time, with a factor of -0.18 mag day$^{-1}$ derived from trial-and-error. After this correction, the secular brightness change within the April epoch appears to be removed, and period searching algorithms (see Section 3.5) suggested a 5.41 hours single-peaked periodicity (Figure 6 middle and lower). However, we note that this purely empirical correction failed to bring the lightcurve from the March epoch to the same magnitude level as the April lightcurve.

Since the aperture photometry, at least those from apertures larger than tens of pixels in radius, is dominated by coma brightness rather than nucleus, the lightcurve amplitude decreases with increasing aperture size. The lightcurve amplitude vanishes at the point where the dust grain aperture crossing timescale is comparable to the time between consecutive lightcurve extrema, which is somewhere between 50 and 60 pixels, or 130-160 km from the nucleus (Figure 7). Using half periodicity of 2.7 hours, the mean expansion speed of the dust in the coma is estimated to be 10-20 m s$^{-1}$. Given the small expansion speed of coma dust, the periodicity in cometary photometry can only be detected when the aperture size is smaller than a few hundred km, which is generally achievable with *HST* but difficult from seeing-limited ground-based observing facilities.

*3.2 Coma color*

From images acquired through the *V*- and *r'*-band, we calculated the linear color slope of the comet. The azimuthally averaged color slope of the dust coma is measured in co-centric annular apertures centered at the nucleus, with the median from all images obtained in the same *HST* orbit plotted (Figure 8). Overall the color measurement has error bars of up to ±2%/100 nm for the March epoch measurements, and up to ±0.3%/100 nm for the April epoch measurements. Although at small distances (<20 pixels or <50 km) from the nucleus, the color measurement could be affected by the smearing, the color plots appear to be smooth and the median appears to have effectively filtered out the potential artifacts due to smearing.

Overall the color slopes of 252P on March 14 and April 4 are consistent with each other, and have a similar cometocentric distance trend. At within 100 km from the nucleus, the color s lope of 252P's dust coma is about (6±1)%/100 nm for both epochs. At between 100 and ~500 km from the nucleus, the color slope decreases (bluing) with distance. At >500-600 km, the dust coma shows a bluer color (negative slope). The decreasing trend in color slope with cometocentric distance on March 14 may be slightly steeper than that on April 4, but not statistically significant. Similarly, any possible variations in the color slope with rotation are not statistically significant in our data.

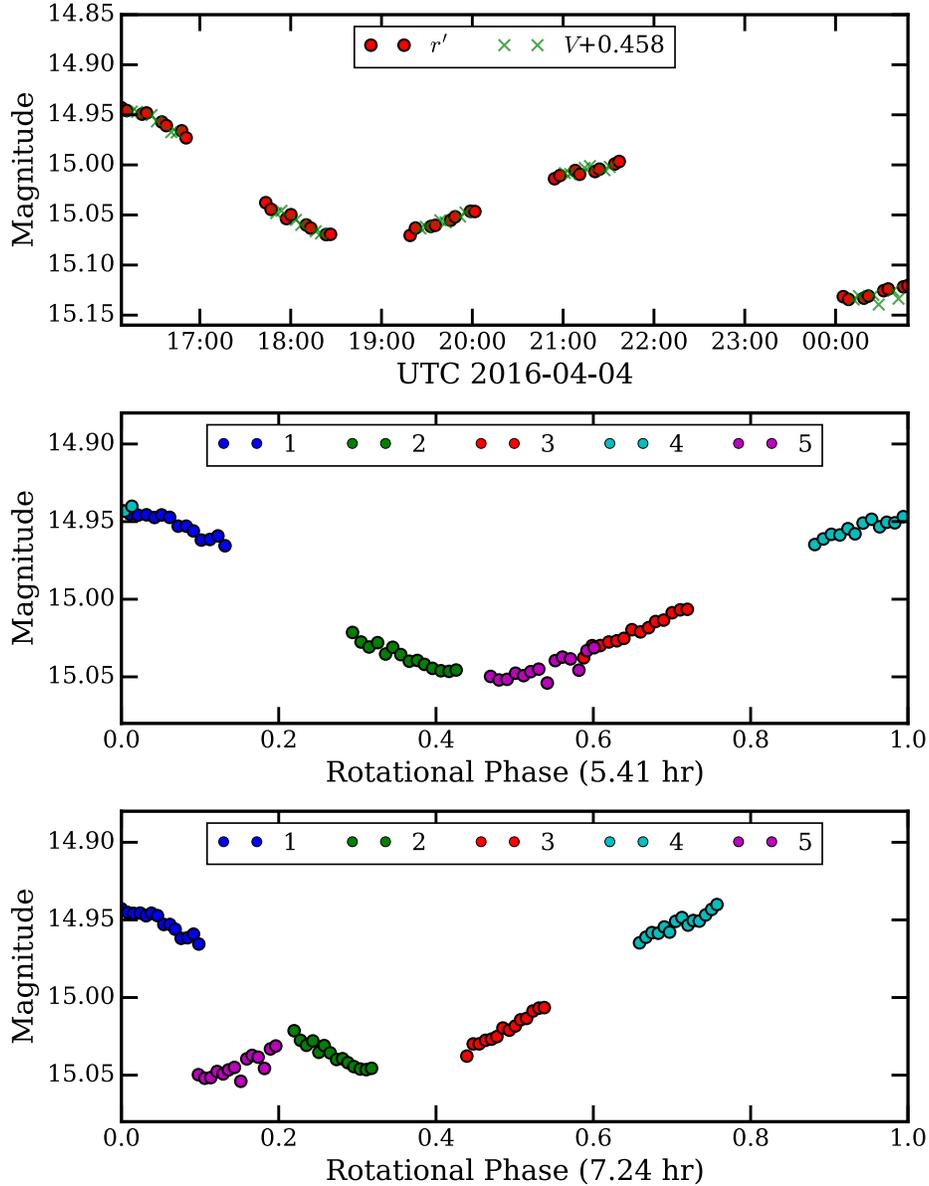

**Figure 6**: Upper panel: The total (Vega) magnitude of 252P in 10-pixel-radius (0.4") aperture from *r'*-band (red circles) and *V*-band (shifted by +0.458 mag) (green cross) measured from images in April epoch. Middle panel: The same data in the upper with the long-term brightness trend removed (see text) and phased by a periodicity of 5.41 hours. Different colors depicts the measurements from different *HST* orbits as noted in the legend. Lower panel: Same as the middle panel, but phased with a periodicity of 7.24 hours. (see online PDF version for color)

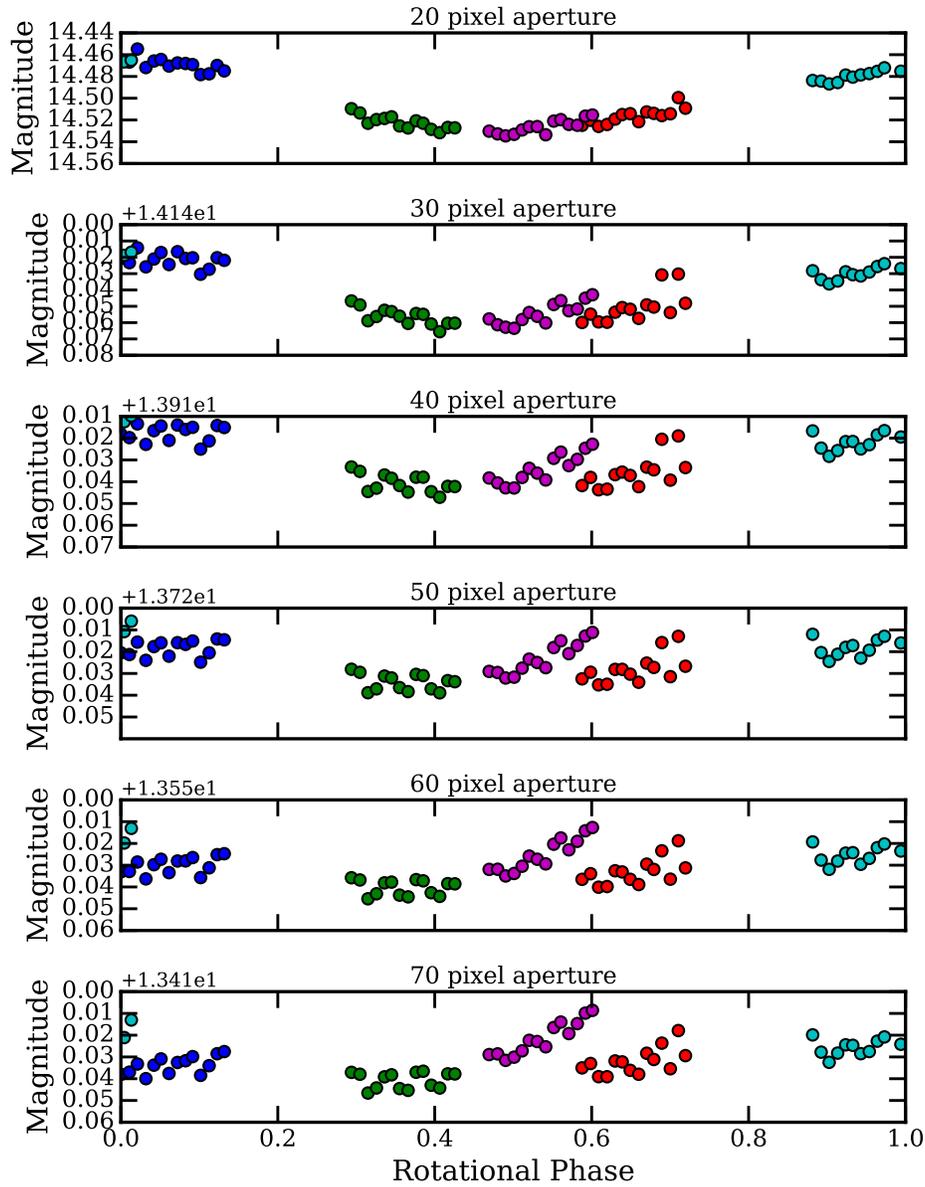

**Figure 7**: The lightcurves of 252P measured from various aperture radii as shown on top of each panel. Different colors of symbols correspond to different HST orbit in the April epoch as in Figure 6. The amplitude of lightcurve decreases with increasing aperture size, and diminishes below the scatter in the data (~0.03 mag) at about 50-60 pixels aperture.

This color-distance behavior of 252P is different from previous *HST* observations of Comets C/2012 S1 (ISON), which showed an increase of color slope (reddening) with distance (Li et al. 2013a), and C/2013 A1 (Siding Spring) had a nearly neutral dependence (Li et al. 2014, 2016), although the trend for those comets are at a distance of 10,000 km from the comet. The factors that could affect the color of dust coma include composition, grain size, and possible contamination from gas emission lines in the filter bandpasses. Comparisons with the typical spectrum of comets (e.g., Feldman et al. 2004) suggest that the *V*-band filter that we used to image the comet covers the Δν=0, 1 bands and part of the Δν=-1 band of $C_2$ gas fluorescent emissions, while the bandpass of the *r'*-filter covers part of the Δν=-1 band and much weaker Δν=-2, -3 bands of $C_2$ gas emission, as well as the weak $NH_2$ bands. Ground-based observations suggest that 252P is extremely gas rich with strong $C_2$ (McKay et al. 2017). From our *DCT* observations (Table 3), the *Afρ* values derived for *BC* and *RC* filters, which are relatively clean of gas contamination compared to *V* and *r'* filters, were 24.8 and 29.7 cm, respectively, whereas the values for *V* and *r'* filters are 39.5 and 34.9 cm, respectively. The higher value from *V*-band suggests more gas contamination than in *r'*-band, consistent with being contaminated by $C_2$. Gas contamination is more evident further out from the nucleus due to the much slower brightness falloff of gas coma compared to the dust, consistent with the color-distance trend that we observed. Therefore, the bluing trend of the coma color that we measured for 252P is likely a result of more gas contamination in the *V*-band filter than in the *r'*-band filter.

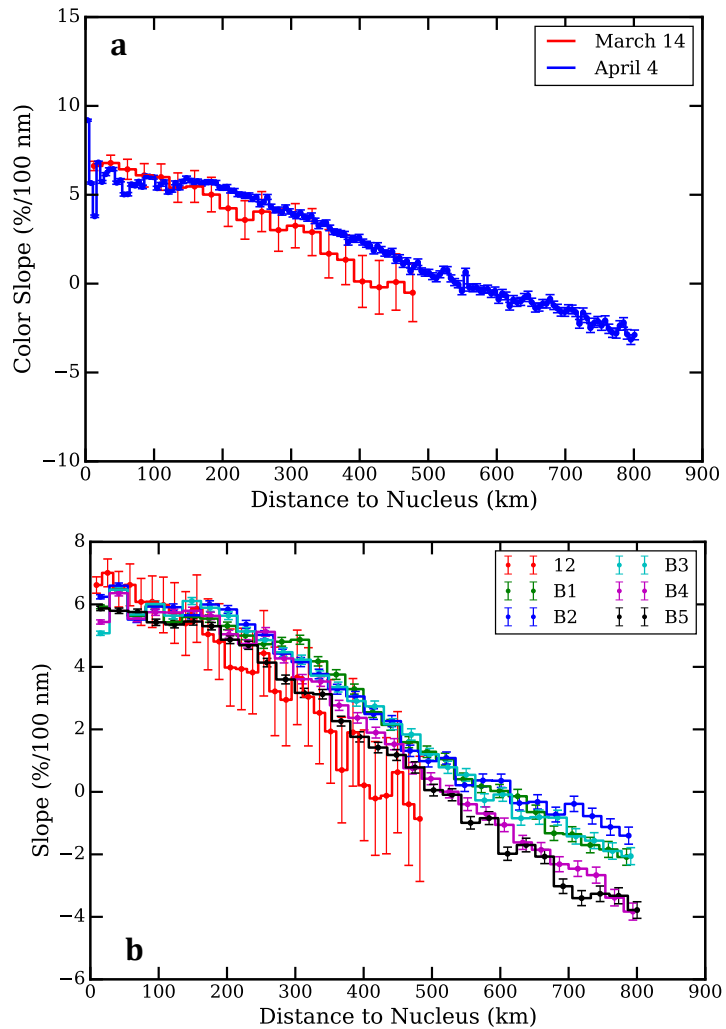

*3.3 Coma morphology*

After enhancing the *HST* images by dividing out the 1/ρ brightness model that represents a coma with a constant dust outflow (e.g., Samarasinha & Larson 2014), we can see obvious jet-like feature in the sunward direction in both epochs of images (Figure 9). The jet feature appears to be narrow and well defined in most images in both epochs, and the length of its brightest core is >20 pixels (~50 km). At 50-100 km from the nucleus, the opening angle of the feature widens to about 30°-40°, diffusing into the background coma at ~200 km from the nucleus. In the April epoch, the projected position angle (PA) of this jet feature changed systematically from ~60º to ~120º, suggesting a rotation of the nucleus over the duration of our observations.

**Figure 8**: Panel a: The color slopes of 252P with respect to distance to the nucleus in the two observing epochs. Panel b: Same as panel a, but with the color measured from each orbit in the April epoch plotted separately.

To estimate the possible periodicity in the PAs of the sunward jet, we measured the PAs of the centerline of the jet feature by fitting a Gaussian and a 4$^{th}$ order polynomial to the azimuthal brightness profile within ±50º of the centerline. The measured PAs from both techniques agree with each other within 5º. The measurements of the PAs at 5-50 pixels from the nucleus showed that the jet feature as projected in the sky plane is mostly radial from the nucleus in the March epoch (Figure 9a, b). In the April epoch when the PAs are <90º (Figure 9c-g), the jet feature is also nearly radial, but slightly curved towards north when PA>90º (Figure 9h-k). Near the maximum PA of ~120º (Figure 9m, n), the jet feature quickly becomes diffuse and non-distinguishable from the background coma. The curvature and the diffuse behavior of the jet feature at this time suggest that it was likely pointing close to and sweeping through the line-of-sig ht (LOS) but at ~120º PA.

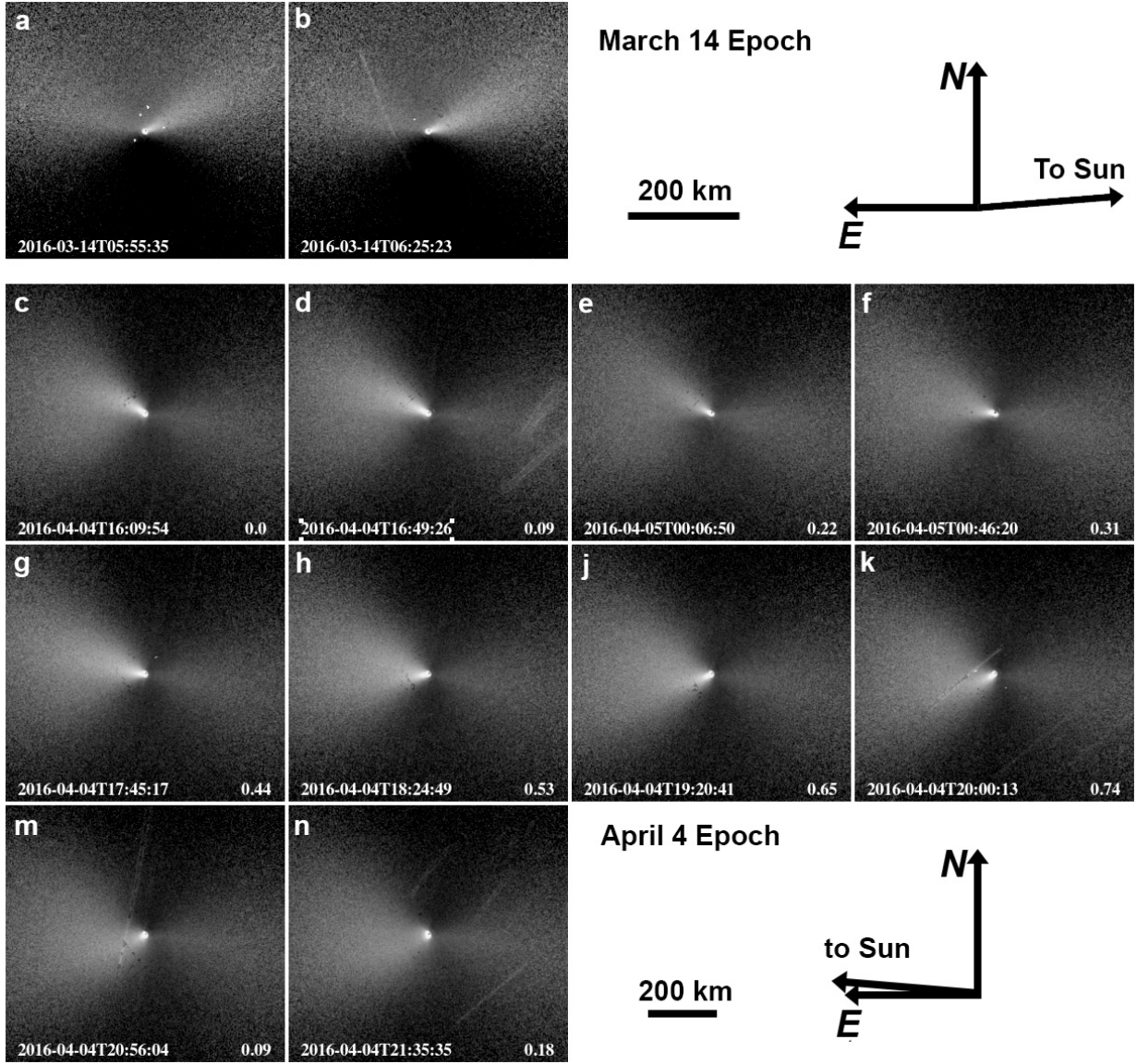

**Figure 9**: The $1/\varrho$ divided images of 252P in *r'*-band taken at the start and the end of each *HST* orbit. Frames a and b are from the March epoch, and other frames are from the April epoch. The frames in the April epoch (c-n) are ordered in their rotational phases for a periodicity of 7.24 hours as marked in the lower right of each panel. The UTCs of each image are in the lower left of each panel. The nucleus is at the center of each frame. All images are north up and east to the left, with the scale bar and the projected orientation of the Sun (265º PA in March and 85º PA in April) marked respectively for the two epochs. Background stars (trails) are removed, although faint star trails are still visible in approximately northwest to southeast direction in some panels.

Figure 10 (upper) shows the measured PA of the jet feature at 11 pixels (30 km) from the nucleus. The scatter in the fourth orbit (cyan symbols) is during a period when the jet was sweeping thro ugh the LOS and became much more diffuse than before, resulting in higher uncertainty in the measured PAs. The PAs measured at other times are all accurate to a few degrees as suggested by the scatter in the data. Since the change in the jet feature should be due to the rotation of the nucleus, we can infer the periodicity from the PA measurements. However, when using the periodicity suggested by the lightcurves, 5.41 hours as discussed earlier, the PAs do not phase together (Figure 10 middle). The best periodicity that can phase the jet morphology data is 7.24 hours (Figure 10 lower). We will discuss the rotation of the nucleus in Section 3.6.

*3.4 Nucleus size*
*3.4.1   Measurements*

The unresolved central portion of each comet image is a combination of scattered light from the nucleus and coma dust, and resonance emission from coma gas. By measuring the inner-coma surface brightness radial profile, we can separate the coma contribution from the nucleus contribution, and estimate the nucleus brightness and size (e.g., Lamy et al. 2011). Our approach is summarized as follows, and is depicted by Figure 11.

We start with images that have had standard calibrations applied (e.g., dark subtraction, flat field correction), but have not had the field distortions removed, preserving the original shape of the point response function. The inner-coma is divided into 30 to 100 azimuthal bins, and a radial power law is fit to the surface brightness profile within a 12 to 100 pixels annulus within each resulting wedge, except for image icvbb3f6q where we used a smaller outer radius of 33 pixels to avoid a background source. The power-law slopes and scales are used to generate a synthetic super-resolution image of a coma (0.1-pixel steps), taking care to estimate the surface brightness on sub-pixel scales. We then convolve the image with a synthetic point source generated with 0.1-pixel steps using the TinyTim software (Krist et al. 2011) version 7.5. Each run of TinyTim considered the filter and array position of the comet from each image, an assumed telescope focus position, and used a G2V spectral energy distribution. A scaled point source image is then added with the same position as the center of the coma, and the image rebinned to the native pixel scale. The image is then convolved with a model of WFC3's charge diffusion, and the result compared to the original data.

The scale factor for the point source is derived from a least-squares fit to the two-dimensional image, and the result saved. We then perform a grid search over a range of sub-pixel positions for the comet, and over a range of telescope focus values. The final comet position and focus value is based on a by-eye examination of the image and fit residuals.

We initially examined images with the least amount of smearing (<1 pixel), but with integrations long enough for a high signal-to-noise detection of the coma (generally 30 s or more). However, the resulting data set was limited. In order to consider images from both March and April, and with both the *V* and *r'* filters, we necessarily included some smeared images. We chose frames with an apparent linear smearing of only a few pixels. The smearing was accounted for by convolving the super-resolution TinyTim point source image with a line. We chose a smear length and position angle based on our estimate of smearing (Section 2.3), but fine-tuned with by-eye examination of the image and fit residuals.

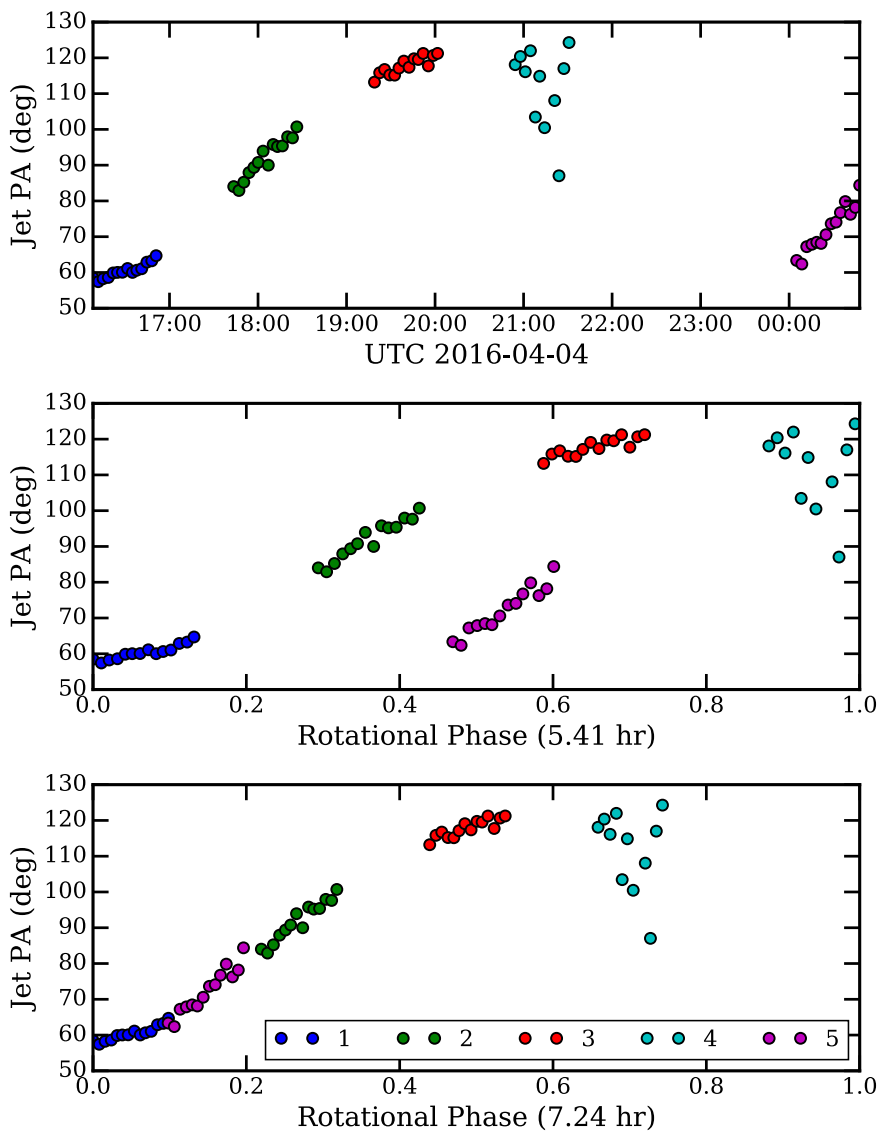

**Figure 10**: Upper panel: The PAs of the jet measured at 11 pixels (30 km) from the nucleus from all images from the April epoch. Different colors are used for the five *HST* orbits. Middle panel: The measured PAs in upper panel phased with a periodicity of 5.41 hours. Lower panel: Same as middle panel but phased with a periodicity of 7.24 hours.

The nucleus size measurements from those six selected images are summarized in Table 4. To convert the total magnitude of the nucleus to its size, we assumed an $r'$-band geometric albedo of 0.04, and a slope of 0.04 mag deg$^{-1}$ for its phase function. The color term $V-r' = 0.463$ mag (~6%/100 nm) in a 3-pixel radius aperture is also taken into account in the conversion from different bands. The smear lengths for images icvb12d4q, icvbb3f6q, and icvb12d1q were 1, 2, and 3 pixels long, respectively. Image icvbb3f6q was from April, and the other two were from March. The agreement between the nucleus size measured from variously smeared images and those not smeared in the same epoch suggest that our approach to account for smearing is robust.

**Table 4**. Summary of nucleus fitting results.

| Image | UT | Band | $r_h$ [1] (au) | $\Delta$ [2] (au) | $\alpha$ [3] (°) | Nucleus | Coma ($10^5$ DN) | Residual | $M_{nuc}$ [4] | $R_{nuc}$ [5] (km) |
|---|---|---|---|---|---|---|---|---|---|---|
| icvb12d1q | 2016-03-14T06:19:23.722 | $V$ | 0.996 | 0.0566 | 86.5 | 0.91 | 1.22 | 0.39 | 17.60 | 0.28 |
| icvb12d4q | 2016-03-14T06:26:40.731 | $r'$ | 0.996 | 0.0566 | 86.5 | 1.09 | 1.30 | 0.21 | 16.94 | 0.31 |
| icvbb1e9q | 2016-04-04T16:28:03.789 | $V$ | 1.037 | 0.0922 | 64.5 | 3.13 | 3.32 | 0.89 | 16.57 | 0.51 |
| icvbb3f5q | 2016-04-04T19:35:31.805 | $r'$ | 1.037 | 0.0929 | 64.4 | 2.70 | 3.16 | 0.57 | 16.27 | 0.48 |
| icvbb3f6q | 2016-04-04T19:38:50.797 | $V$ | 1.037 | 0.0929 | 64.3 | 3.11 | 3.19 | 0.31 | 16.58 | 0.51 |
| icvbb4ftq | 2016-04-04T21:01:08.797 | $V$ | 1.037 | 0.0933 | 64.3 | 3.98 | 2.93 | 0.25 | 16.31 | 0.58 |

Notes:
[1] Heliocentric distance
[2] Geocentric distance
[3] Solar phase angle
[4] Nucleus magnitude
[5] Nucleus radius

*3.4.2 Quality assessment*

The quality of the nucleus extraction for 252P can be assessed by comparing the total counts of nucleus, coma, and residual from the extraction (Table 4), inspecting the residual images after subtracting the nucleus model and coma model from the original image (Figure 11), and comparing the nucleus size measurements from the two epochs.

For all six images, the sums of the residuals are all positive at 4-18% level of the total flux, and the standard deviations of the residuals are <7% of the residual, suggesting that some coma signal is indeed missing from the coma model. As shown in Figure 11, the positive residual is mostly symmetric about the nucleus, with some subtle enhancement towards where the strong sunward jet is located. Based on these morphological characteristics of the residual, we suggest that there are two possible sources for the coma: the jet that is asymmetric with respect to the nucleus, and the enhanced dust coma that is symmetric near the nucleus.

In almost all images of 252P, the bright jet is visible in the sunward direction, with its bright core close to the nucleus and outside of the region where we fit the coma. Thus, the jet cannot be fully included in the coma model once extrapolated into the center region. Since we fit the nucleus with a PSF, and use the best-fit PSF to measure the nucleus brightness, any asymmetric residual coma structure in the center region is mostly excluded from the nucleus measurement. Therefore, the photometric measurement of the nucleus based on this technique should be robust against most asymmetric coma residual.

The symmetric residual pattern is somehow puzzling. It is not likely dominated by the inaccurate PSF, because a test of the nucleus extraction with the synthetic PSFs at various focus suggested that the PSF we adopted resulted in the lowest overall residual. A narrower PSF would result in a wider residual ring, while a wider PSF would result in a bright residual spot in the center, both with an overall higher residual. One potential physical interpretation is that such a symmetric residual pattern represents enhanced coma dust in near nucleus environment. To investigate the possibility of this explanation, we compared the nucleus size measurement from the images in the same epoch and from two different epochs in March and April.

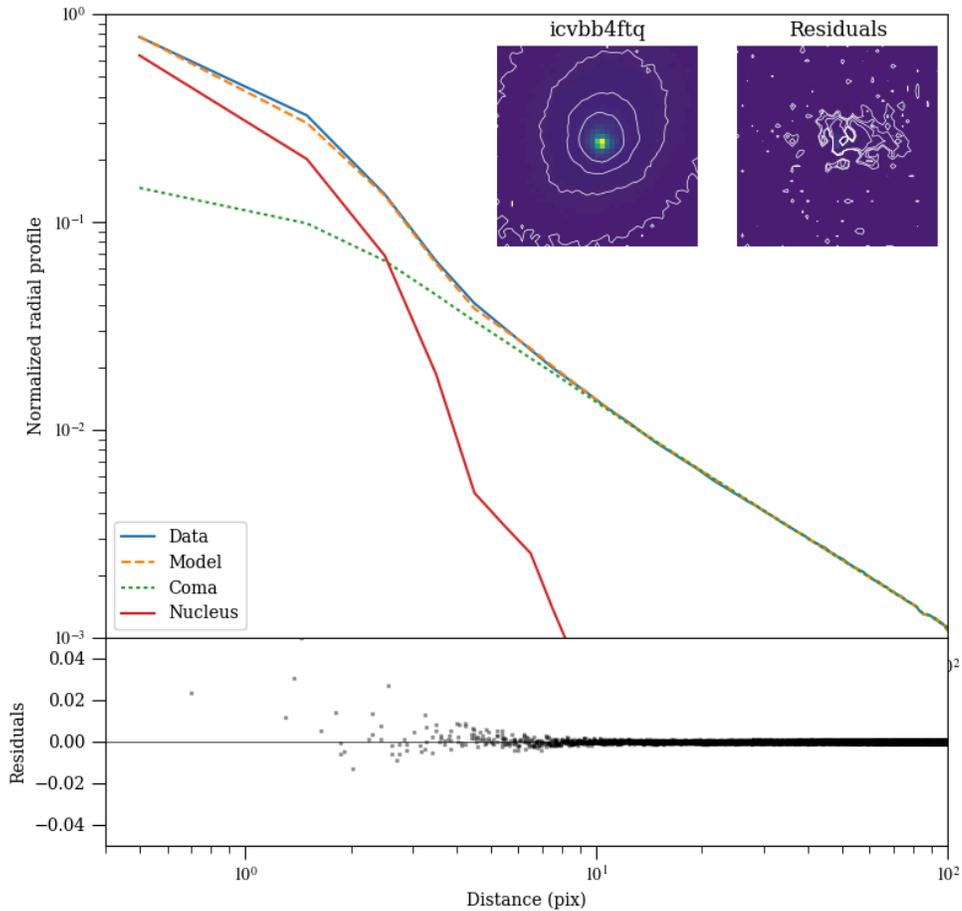

**Figure 11**: The nucleus extraction results for image **icvbb4ftq**. The contours in the left (original) image are 4%, 2%, 1%, and 0.5% of the coma peak; and the contours in the right (residual) image are 0.4%, 0.2%, 0.1%, and 0.05% of the original coma peak. Overall the model fits the data well except for some positive residuals within a few pixels from the nucleus.

The nucleus radius measured from the two images in the March epoch are consistent with each other at 0.30 km, and those from the April epoch are within 11% of 0.52 km, despite being measured from two different filters in both epochs, demonstrating that the measurement is quite stable even though a considerable level of residual coma is not included in the coma model, and various amount of smearing appears in the images. Note that an upper limit on nucleus radius of 252P can be derived using the MPC photometric data near the start (September to November 2015) and the end (July 2016) of its 2016 apparition, when the comet was very weakly active or inactive and its total brightness appears to follow $1/\Delta^2$ scaling, where $\Delta$ is geocentric distance. We find a limit between 0.35 and 0.5 km, consistent with our measurement, and the difference could suggest that we may still have some residual coma in the nucleus measurements, or are observing different aspects of a nucleus with an elongated shape.

On the other hand, the nucleus size measurements from two epochs differ by nearly 60%, representing a difference of a factor of ~3 in the total brightness and cross-section. Although the derived nucleus size depends on the assumed phase slope, in order to fully account for the difference in the nucleus size measurement from the two epochs separated by 22° in phase angle, a phase slope of 0.09 mag deg$^{-1}$ is required, which is considered implausible (e.g., Li et al. 2013b). Even for a phase slope as steep as 0.06 mag deg$^{-1}$, which is steeper than that of any cometary nuclei reported before, the nucleus size from the two epochs still differ by ~40%, and total cross-section a factor of ~2. Therefore, the assumption of phase slope cannot explain such a large difference in derived nucleus size from two epochs.

Here we consider two possible explanations. The first is that we observed a highly elongated nucleus from two different viewing aspects separated by 120° from the March epoch to the April epoch. Since the April epoch cover nearly one full rotation and our measured nucleus sizes differ by just 20%, either the nucleus is nearly pole-on in this epoch, or it is not elongated. If the former is true, then we need to be exactly looking along the long-axis of the nucleus in our March observations in order to measure its smallest cross-section. We consider such a probability is low, and the first explanation is not likely true although not entirely impossible.

The second explanation has to do with the residual coma within the PSF that cannot be removed by the coma model but accounted for by the PSF fitting. Such near nucleus coma could also be the source of the symmetric residual pattern that we discussed earlier. Rosetta observations from close distance of 67P suggested a cloud of large grains that appears to be encircling the nucleus of 67P in bound orbits (Rotundi et al. 2015, Davidsson et al. 2015), similar to some extent to the centimeter to decimeter-sized icy chunks around the nucleus of 103P/Hartley 2 (hereafter 103P) (Kelley et al. 2013, Hermalyn et al. 2013). The Hill sphere of 252P is about 30-50 km, assuming the typical density of 500 kg m$^{-3}$ for cometary nuclei. Considering that the inner 1/3 of Hill sphere is more likely to host stable orbits of bound grains (Toth 1999, Tricarico & Samarasinha 2011), and the pixel size of 1.6 and 2.7 km at the nucleus in March and April epochs, respectively, then if such bound grains also exist around Comet 252P, they would be most likely within 3-10 pixels from the nucleus in our images, and could not be fully included in the coma model. In this case, they would both result in a symmetric residual and leave some signal in the fitted PSF that could be mistakenly considered as from the nucleus. The 1.6× larger pixel scale of the April images over that of the March images means that it is harder to include those grains in the coma model for April images, resulting in an apparent increase in the measured nucleus brightness. In addition, if the increased activity of 252P from March to April causes an increase of the near-nucleus grains, then it could also increase the measured nucleus size. In this explanation, the nucleus size measurement from the April epoch would be an upper limit, and the March measurement would be more reliable. In this case, the nucleus radius is close to 0.3 km, and we estimate an error bar of about 10%, or 0.03 km.

For a quantitative assessment of the near-nucleus dust scenario, if the size distribution for bound grains is assumed to be a power law with an exponential of -3.5 within an upper and a lower size limit, for a given total cross-section, the total mass is derived to be proportional to the geometric mean of the largest and smallest grains. If we further assume a density of 1000 kg m$^{-3}$ for the grains, and a geometric mean size of 1 mm as suggested by 67P and 103P results, then the total mass of the grains increased from March to April would be about 5×10$^5$ kg. This mass is about 0.001% of the total mass of a 0.3 km radius nucleus (assuming a density of 500 kg m$^{-3}$), or a layer of ~1 mm in depth on the surface, which is completely plausible.

*3.5 Active area*

The OH production rate of 252P on April 17, 2016 is $Q$(OH) = 5.8×10$^{27}$ s$^{-1}$ from our *DCT* data (Table 2). This value is comparable to that measured on in late March by TRAPPIST (4.5±1.0×10$^{27}$ s$^{-1}$; E. Jehin & C. Opitom priv. comm.), and 10× of that in late-February (3.9×10$^{26}$ s$^{-1}$; D. Schleicher priv. comm.), consistent with each other considering the heliocentric distance trend of the sublimation activity.

Based on the sublimation model of Cowan & A'Hearn (1984), the sublimation rate of water ice at 1 au is about 1.6×10$^{22}$ s$^{-1}$ m$^{-2}$ for a slow rotator model, and 3.3×10$^{21}$ s$^{-1}$ m$^{-2}$ for an isothermal model. Assuming a branching ratio of 0.86 from water to OH by photodissociation (Huebner et al. 1992), the $Q(H_2O)$ of 252P in late April 2017 is about 7×10$^{27}$ s$^{-1}$, and requires an active area of 0.4 and 1.8 km$^2$ on the nucleus for the two extreme models, respectively. For a nucleus of 0.3 km radius, the active fraction of 252P is from ~40% to over 100%.

For typical JFCs, the active area is a few percent (A'Hearn et al. 1995, Lamy et al. 2004). Some small comets, such as 103P (Groussin et al. 2004) and 46P/Wirtanen (Lamy et al. 1999), have active fraction of >100% near perihelion. Spacecraft images from close distances of 103P not only showed that its gas activity still appear to originate from discrete areas on the nucleus (Protopapa et al. 2014), but also resolved numerous icy chunks centimeters to decimeters in size ejected from the nucleus and sublimating as they move away (Kelley et al. 2013, Hermalyn et al. 2013), increasing the total active area of the comet as measured remotely. Thus, the large active fraction of 252P of possibly near or greater than 100% could also indicate a large amount of icy grains surrounding the nucleus, consistent with the scenario we proposed previously (Section 3.4). Considering that the dust activity level of 252P was much weaker than this current 2016 apparition, then those icy grains, if indeed there, should be ejected in this current apparition rather than in the past.

*3.6 Rotation of the nucleus*

We searched for periodicities in both the photometry data from 10-pixel radius apertures, and the jet PA data at 11 pixels from the nucleus with two different period searching algorithms, the Lomb-Scargle (LS) method (Lomb 1976, Scargle 1982) with super smoother, and the Bayesian generalized Lomb-Scargle (BGLS) method (Mortier et al. 2015, Zechmeister & Kürster 2009). The LS periodogram expresses the period search results in the arbitrary power, and the periods with the highest power are considered the most likely periods in the data. The BGLS periodogram is extended from the LS method by using the Bayesian probability theory to describe the probability in frequency/period space. Figure 12 shows the probability plot of our period search with the *V*-band photometric data using the BGLS method as an example. We also searched the period in *r'*-band photometric data and the combined dataset with both *r'*-band and *V*-band (shifted by the *V-r'* color of 0.458 mag). All methods for all photometric datasets returned a similar periodicity of 5.41±0.07 hours, and for the jet PA data returned a periodicity of 7.24±0.07 hours. The uncertainties in the best-fit periods are estimated as the standard deviation of the approximate Gaussian shape of the power-frequency plots (for LS method) and the probability frequency plots (for BGLS method). The best-fit periodicities for both the lightcurve and the jet PA do result in good phasing of data (Figures 6 & 10).

Comet 252P was observed from the ground before and after its Earth encounter, and the observing conditions during its encounter were unfavorable for most ground-based telescopes due to its southern declination and the full Moon. Knight & Schleicher (2016) reported observations from the ground in early April 2016 in dust, CN, and OH. They detected a short sunward dust feature that has a length and projected orientations consistent with our April epoch observations. On the other hand, the gas coma in both CN and OH displays a different morphology from the dust coma (Figure 3). In particular, what appeared to be two coma features in the gas were visible in the sunward side of the nucleus, and varied smoothly along the north-south directions. Due to their vastly different size scales, it is unclear whether the dust feature and one gas feature originate from the same sources on the nucleus. Despite this, the CN coma morphology as they observed varied smoothly during a night and showed repeatability in ~22 hours and ~95.5 hours, implying a period of 7.35±0.05 hours. This periodicity is consistent with the value that we derived from dust morphology, but not with the value derived from our lightcurve data.

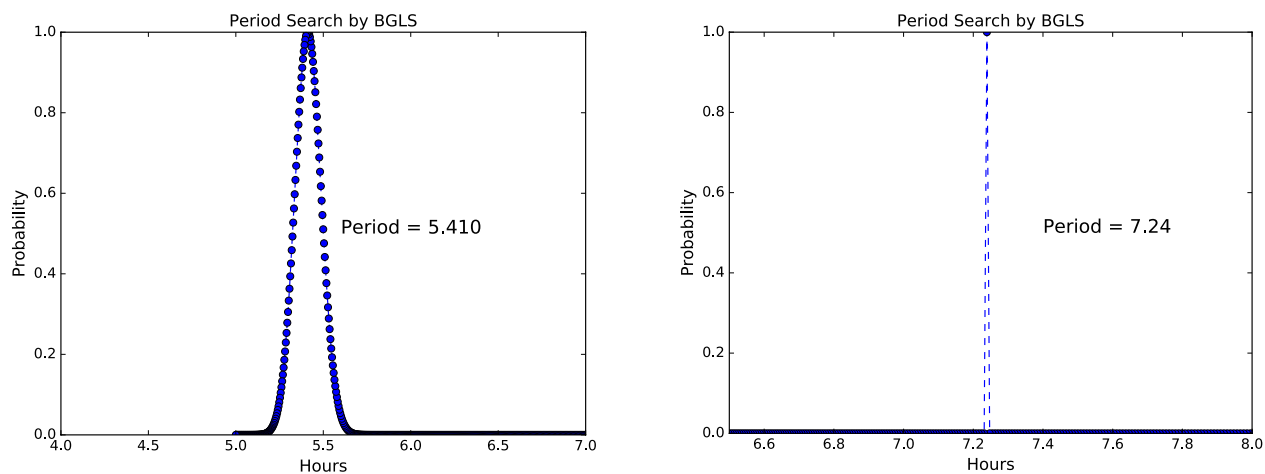

**Figure 12**: The BGLS probability plot for period search in the *V*-band photometry data (panel a) and jet PA data (panel b). The peak probability is normalized to unity.

In order to assess the relative contributions of the sunward jet and the ambient coma to our photometric lightcurve, we divided the circular aperture into two half-circles, one centered at the azimuthal direction of the jet in the sky, and the other centered at the anti-jet direction. The azimuthal profiles of the coma showed much larger azimuthal variations in the jet side than the anti-jet side (Figure 13a). Therefore, we use the total brightness on the anti-jet side to approximate the total brightness of the ambient coma on the jet side, and subtract it from the jet-side photometry to estimate the fractional contribution of jet in the coma lightcurve (Figure 13b). Overall, the sunward jet contributes a few percent to up to 20% of the total brightness in the 10-pixel-radius aperture photometry. In addition, the 5.41 hours periodicity keeps visible in full-circle aperture lightcurves for larger apertures until about 50 pixels radius aperture (Section 3.1, Figure 7), consistent with being dominated by the overall dust coma rather than the jet. Therefore, we conclude that the jet and the ambient coma have different sources on the nucleus, and the lightcurve and the associated 5.41 hours periodicity (Figure 6) are dominated by the ambient coma rather than the jet.

Details of how cometary jets hundreds to thousands of km in extent form and extend to space are largely unknown. Recent spacecraft observations of comets from close distances revealed a large number of small jets that only extend a few to tens of km from the nuclei, and some attempts have been made to correlate those small jets to large-scale jets observed from the ground (e.g., Farnham et al. 2007, 2013). In the case of 19P/Borrelly, the large-scale sunward jet observed from the ground directly originates from the nucleus (Soderblom et al. 2004). In other cases, at least some small jets near the nucleus probably merge together at some distance from the nucleus to form large-scale jets that are observable from ground (e.g., Sekanina et al. 2004). On the other hand, at least for the case of 67P, the observations by Rosetta suggested that the whole sunlit areas together with multiple sources that produce enhanced dust release are the dominant source of its ambient coma (Kramer & Noack 2016). If this is the case for 252P, then the modulation of the whole sunlit area due to the rotation of the nucleus could cause the modulation of the dust release into ambient coma, while the large-scale jet that we observed could originate from a relatively small, isolated area on the nucleus. Therefore, the rotational modulations of the lightcurve and that of the projected PA of the jet could be different for 252P: the lightcurve reflects the change in the entire sunlit area, while the jet reflects the illumination condition and the orientation of its source region.

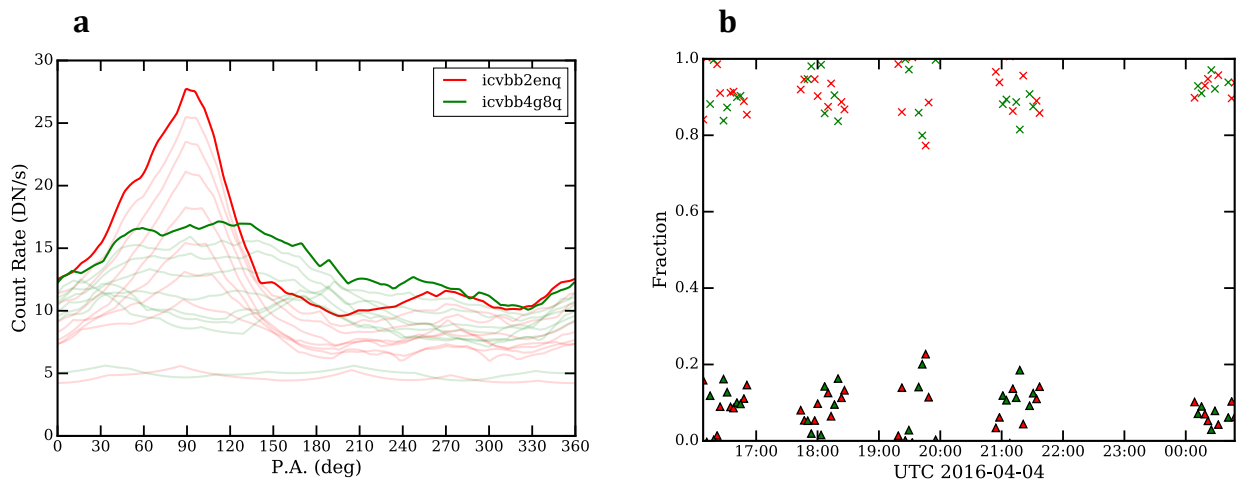

**Figure 13**: Panel a: The azimuthal profiles of two *r'*-band images. Image **icvbb2enq** (red lines) corresponds to the middle of the 2$^{nd}$ *HST* orbit in the April epoch, when the jet appears to be sharp and well defined, and straight out from the nucleus. Image **icvbb4g8q** (green lines) corresponds to the end of the fourth *HST* orbit in the April epoch, when the jet appears to be mostly diffuse due to foreshortening. The dark solid lines are the profiles at 10 pixels from the nucleus, while the shaded lines are the profiles at progressively smaller distances from the nucleus inversely scaled by the cometocentric distance squared. Panel b: The estimated fractional contribution of the total brightness in a 10-pixel radius aperture from the jet (triangles) and from the ambient coma (crosses). Red and green symbols mark the measurement from *r'*-band and *V*-band, respectively.

## 4. Discussion
*4.1 Non-principal axis rotation of 252P*

The different periodicities derived based on the repetition of coma morphology and lightcurves present a dilemma. If the nucleus is in the most dynamically stable rotation around the short principal axis, under certain circumstances, periodicities in the 1:2 ratio can be understood. However, a periodicity ratio of 4:3 (7.24 hours vs. 5.41 hours) cannot be explained in terms of a principal axis rotator, and we suggest that Comet 252P is likely in a non-principal-axis (NPA) rotational state. The limited amount data on this comet makes it difficult to derive a definite rotational state. However, three additional lines of evidence support the possibility of an NPA rotator.

First, since the nucleus of 252P is relatively small, it is a likely candidate for rapid changes in its rotational state including the possibility for excitation into an NPA rotational state (cf. Samarasinha & Mueller 2013 and references therein). NPA rotational states are not rare in comets, with best examples being Comets 1P/Halley (e.g., Samarasinha & A'Hearn 1991) and 103P (Belton et al. 2013). Second, numerical studies indicate that as rotational states of comets evolve, they may spend considerable amounts of time near states having commensurable component periods (Samarasinha & Belton 1995, Neishtadt et al. 2002). This is also the case for 1P and 103P. Third, the activity of 252P in its 2016 apparition increased by a factor of nearly 100 from its 2000 apparition level (Section 4.3). If we fit the non-gravitational force to 252P's orbit before and after April 1, 2016 separately, the amplitudes of $A_1$ and $A_2$ parameters both increase by a factor of ~10. Such a sudden and significant increase in the activity level on a small nucleus could likely change the rotational state and even put it into excited rotation during its 2016 perihelion passage.

We can propose one possible scenario for the rotational state of 252P to explain the conflicting periodicities derived from coma morphology and lightcurve. If the partial lightcurve-based periodicity (5.41 hours) is manifesting the total outgassing from the entire sunlit side of the nucleus, it may correspond to a period $P$=5.41 hours given by $\frac{1}{P} = \frac{1}{P_\phi} + \frac{1}{P_\psi}$ where $P_\phi$ is the precessional period of the long axis (7.24 hours) and $P_\psi$ is the rotational period of the long axis around itself (~22.0 hours). If the morphological features in the dust (and in *CN*) are originating from a source region near the end of the long axis, it may only be mimicking $P_\phi$. One potential difficulty for this scenario, though, is that if there are two CN jets observed from the ground and both rotating about the nucleus at a period of 7.35 hours (Knight & Schleicher 2016), then they would have to both originate from close to the two ends of the long-axis. More data from the ground are needed to study the phasing of the two CN jets and see how they are related to each other. While this scenario may not be the only solution, it is a possible scenario.

*4.2 Comparisons between 252P and BA$_{14}$*

BA$_{14}$ had a close encounter to Earth at 0.0237 au on March 22.64682, 2016, one day after the close approach of 252P. The total magnitude data of BA$_{14}$ from the MPC can be fitted with a scaling of $(r_h \Delta)^{-2}$ and the typical phase slope between 0.02 and 0.06 mag deg$^{-1}$ for cometary nuclei and dark asteroids (e.g., Lamy et al. 2004, Li et al. 2013b) (Figure 14). In addition, the non-gravitational force parameters $A_1$, $A_2$, and $A_3$ for BA$_{14}$ are constrained to be <10$^{-10}$ au d$^{-2}$ in their absolute values, which are one or two orders of magnitude lower than typical comets (Farnocchia et al. 2014) and consistent with lack of non-gravitational forces. Therefore, we assume that the total brightness of BA$_{14}$ throughout its perihelion passage has always been dominated by the nucleus, and derived an absolute magnitude H=19.2 for BA$_{14}$ with a phase slope of 0.04 mag deg$^{-1}$. The corresponding diameter is ~1 km if we further assume a geometric albedo of 4%. This estimate is consistent with the results from radar observations, which has a diameter of >1 km (Naidu et al. 2016), and the estimate from NEOWISE data (700 m, J. Bauer priv. comm.) and ground-based observations with IRTF (V. Reddy priv. comm.).

Despite having about 10× surface area of that of 252P, BA$_{14}$ has much weaker activity than 252P. We observed the two comets by the *DCT* at similar heliocentric distances. Thus, we can use their CN production rates as a proxy for water, assuming the CN/OH coma mixing ratio is constant. Thus, comet 252P was ~900× more active than BA14, and the active fraction of BA$_{14}$ is about 10$^{-4}$ of that of 252P at ~0.01% level.

Radar observation also revealed BA$_{14}$ as a slow rotator with a period of ~40 hr, (Naidu et al. 2016), which is much slower than that of 252P. The limited duration of radar data cannot constrain the mode of its rotation. Given the relatively large size of BA$_{14}$, it is probably in a relaxed rotational state with any NPA mode damped out already. On the other hand, much smaller size of 252P compared to that of BA$_{14}$ and the active nature of 252P suggest that it is more likely for 252P to have changed its rotational state.

Whether 252P and BA$_{14}$ are a split pair is still unclear. The similarity in their orbits and that in their cometary activities (excluding the elevated activity of 252P in the 2016 apparition) suggest that they are associated. Split events are not rare among JFCs. 169P/2002 EX12 (LINEAR) (hereafter 169P) and P/2003 T$_{12}$ (SOHO) are another example of such a pair (Sosa & Fernández 2015). Comet 289P/Blanpain (D/1819 W$_1$) (hereafter 289P) is a small (300-400 m diameter) and very weakly active JFC that survived from a split event (Jewitt 2006). No other asteroidal or cometary object that appears to be dynamically associated with 289P, but a meteoroid stream is confirmed to be associated with it (Jenniskens & Lyytinen 2005).

*4.3 Original nucleus size*

A 0.3 km nucleus makes Comet 252P one of the smallest cometary nuclei known to date (Fernández et al. 2013). Comet 252P appears to be hyper active in its 2016 apparition with nearly 100% or even higher active fraction. However, the dust activity in its 2016 apparition is substantially stronger than that in its 2000 apparition (Figure 15), which had similar observing geometry to its 2016 apparition. The $Af\rho$ measured from the 2000 apparition of about 0.6 cm (30-90 days post-perihelion), together with the lack of evidence for a dust trail associated with this comet suggest that its activity has remained weak since entering its current orbit a few hundred years ago (Ye et al. 2016a). Although observations during its 2005 and 2010 apparitions are unavailable due to the unfavorable geometries, the success in fitting its orbit solution using the astrometric data from both the 2000 apparition and 2016 pre-perihelion (Section 2.1) suggests that the activity level of 252P in the previous two apparitions was not unusual.

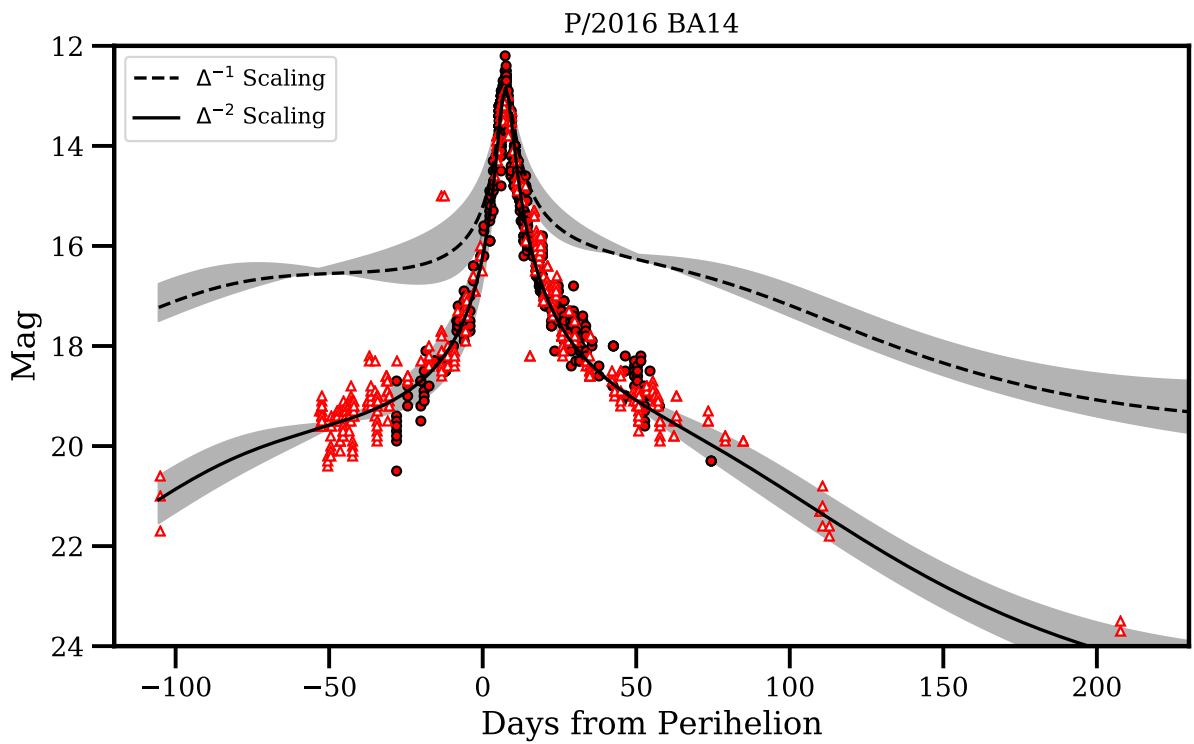

**Figure 14**: The magnitude of $BA_{14}$ from the MPC database. Filled circles are for "total" magnitude, and open triangles are for "nucleus" magnitude. The difference between total and nucleus magnitudes are small, due to the extremely weak activity of the comet. The solid and dashed lines are for magnitude models using different scaling laws with respect to geocentric distance $\Delta$ after the $r_h^{-2}$ and 0.04 mag deg$^{-1}$ phase function correction. Coma brightness is expected to be close to $\Delta^{-1}$ scaling, while nucleus brightness is expected to follow $\Delta^{-2}$ scaling. The shades mark the ranges of magnitudes using phase slopes between 0.02 and 0.06 mag deg$^{-1}$, with the absolute magnitude adjusted accordingly to match data. The apparent brightness of the nucleus is not sensitive to phase slope within the typical range for cometary nuclei.

In the 2016 apparition, the activity level of 252P remained similar to its 2000 apparition level until about two weeks pre-perihelion when its brightness increased significantly. Such an elevated dust activity gradually decreased, and returned to a level comparable to its 2000 apparition level at about 70 days post-perihelion. The sudden increase in non-gravitational force (Section 4.1) during its 2016 perihelion passage strongly suggests that the significant increase in activity level started from this apparition. The reason for the elevated dust activity is not clear. The $Af\rho$ of 50-60 cm about 20 days post-perihelion measured from our data is nearly 100× higher than its activity level in the 2000 apparition at comparable heliocentric distance, and represents the peak dust activity in this 2016 apparition. If its dust/gas ratio remains constant, then the active fraction of 252P in the 2000 apparition would be just <1%, putting it into the category of weakly active comets.

If the increased activity level of 252P in this apparition is unusual, and the activity level of this comet has remained weak as suggested by previous studies, then the small size of this comet could be primordial. However, the possible dynamic link between 252P and $BA_{14}$ means that if they are indeed split pairs, then its size will not be primordial. In this case, if we assume that these two objects are the dominant components of all fragments, then the diameter of the parent comet would be about 1-2 km, which would still be near the small end of the cometary nucleus size distribution (Fernández et al. 2013). The weak active fraction of both objects suggests small primordial sizes of their parents likely, whether a common one or not.

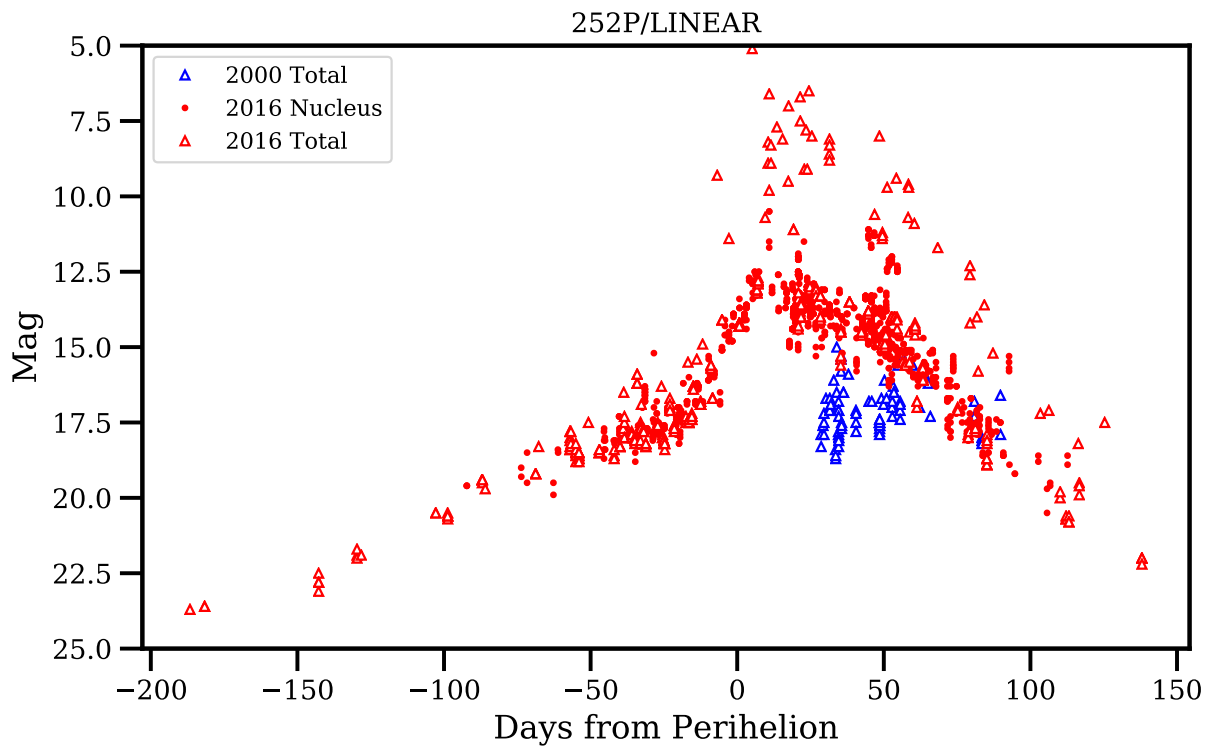

**Figure 15**: The magnitude of Comet 252P from the MPC database. No data is available from the 2005 and 2010 apparitions. Symbols marked as "Total" are the total nucleus and coma magnitude and symbols marked as "Nucleus" are nucleus magnitude, as designated by MPC. No nucleus magnitude data is available for the 2000 apparition.

*4.4 The end state of JFCs*

The weak activity and the possible association of 252P and $BA_{14}$ offer some hints about the end state of JFCs. The possible end states of JFCs include dormancy, extinction, fragmentation, impact on a planet or the Sun, and removal from inner solar system due to planet encounters (mainly with Jupiter). It is not clear how the dormancy and/or extinction occur – whether it is a gradual process and the comet activity decreases slowly, or whether it is a relatively abrupt process when the (near-surface) volatiles in a cometary nucleus depletes. The existence of a number of extremely weakly active comets, including 169P (Kasuga et al. 2011), 209P/2004 CB (LINEAR) (Ye et al. 2016b), 252P, 289P (Jewitt 2006), 300P/2005 JQ5 (Catalina) (Ye et al. 2016b) suggest that the shutoff process is probably gradual.

On the other hand, although the causes for cometary nucleus fragmentation are not obvious and not well understood (e.g., Boehnhardt 2004), most comets that we have observed fragmentation events have moderate to high activity level, although observational selection effect could have played a role. The fragmentation events of those weakly active comets, including 289P parent comet (Jenniskens & Lyytinen 2005), and the parent comets of the putative pairs of 169P and P/2003 T12, as well as 252P and $BA_{14}$ indicate that fragmentations also occur in weakly active comets quite frequently, whether on their way to dormancy or extinction or being weakly active ever since their formation. Therefore, fragmentation and dormancy/extinction are not necessarily exclusive of each other. A common pattern for the end state of JFCs might be a long-term, gradual decrease in cometary activity towards dormancy or extinction, accompanied by occasional fragmentation events. One could imagine that the parent of 252P and $BA_{14}$ might behave like 252P. Although it is weakly active during most of its lifetime, occasional significant increase in the activity might trigger a fragmentation event, ejecting an inert section of its nucleus, and such processes just keep going on. If this is the dominant pattern for the end state of JFCs, then the implication is that most dormant/extinct comets would be fragments and have relatively small sizes, probably around km in diameter.

*Acknowledgement*: This research is supported by NASA through grant HST-GO-14103 from the STScI. The authors acknowledge C.E. Woodward, Y. Wang, Q. Yang, X. Zhou, S. Greenstreet, E. Gomez, A. Heinze, Y. Ramanjooloo, D. Hung, M.M. Knight, L.M. Feaga, T.L. Farnham, S. Protopapa, M. Fausnaugh, R.M. Wagner, D. Koschny, M. Busch, A. Knöfel, E. Schwab, and P. Veres for providing astrometric data of 252P in preparation for the HST observations. D. Farnocchia conducted this research at the Jet Propulsion Laboratory, California Institute of Technology, under a contract with NASA. These results made use of the Discovery Channel Telescope at Lowell Observatory. Lowell Observatory is a private, non-profit institution dedicated to astrophysical research and public appreciation of astronomy and operates the DCT

**References**

A'Hearn, M.F., Schleicher, D.G., Feldman, P.D., et al. 1984, AJ, 89, 579
A'Hearn, M.F., Millis, R.L., Schleicher, D.G., et al. 1995, Icar, 118, 223
Astropy Collaboration, Robitaille, T. P., Tollerud, E. J., et al. 2013, A&A, 558, A33
Belton, M.J.S., Thomas, P., Li, J.-Y., et al. 2013, Icar, 222, 595
Boehnhardt, H. 2004, In: Comets II, pp. 301
Chesley S.R. & Yeomans D.K. 2005, IAU Colloq., 197, 289
Chesley S.R. & Yeomans D.K. 2006, Adv. Astronaut. Sci., 2, 1271
Cochran, A.L. & Schleicher, D.G. 1993, Icar, 105, 235
Cowan, J.J., A'Hearn, M.F. 1984, M&P, 21, 155
Davidsson, B.J.R., Gutiérrez, P.J., Sierks, H., et al., 2015, A&A, 583, A16
Deustua, S., ed. 2016, "WFC3 Data Handbook", Version 3.0 (Baltimore: STScI)
Deustua, S.E., Mack, J., Bowers, A.S., et al. 2016, WFC3 InSR 2016-03 (Baltimore: STScI)
Farnham, T.L., Schleicher, D.G., & A'Hearn, M.F. 2000, Icar, 147, 180
Farnham, T.L., Wellnitz, D.D., Hampton, D.L., et al. 2007, Icar, 187, 26
Farnham, T.L., Bodewits, D., Li, J.-Y., et al. 2013, Icar, 222, 540
Farnocchia, D., Chesley, S.R., Chodas, P.W., et al. 2014, ApJ, 790, 114 (7pp)
Farnocchia D., Chesley, S.R., Micheli, M., et al. 2016, Icar, 266, 279
Farnocchia, D., Bodewits, D., Kelley, M.S.P. et al. 2017, IAU ACM meeting
Feldman, P.D., Cochran, A.L., Combi, M.R. 2004, In: Comets II, 425
Fernández, Y.R., Kelley, M.S., Lamy, P.L., et al. 2013, Icar, 226, 1138
Groussin, O., Lamy, P., Jorda, L., et al. 2004, A&A, 419, 375
Groussin, O., Sierks, H., Barbieri, C., et al. 2015, A&A, 583, A36
Haser, L. 1957, BSRSL, 43, 740
Hermalyn, B., Farnham, T.L., Collins, S.M., et al. 2013, Icar, 222, 625
Huebner, W.F., Keady, J.J., & Lyon, S.P. 1992, Ap&SS, 195, 1
Hunter, J.D., 2007, Computing in Science & Engineering, 9, 90
Jenniskens, P., Lyytinen, E. 2005, AJ, 130, 1286
Jewitt, D. 2006, AJ, 131, 2327
Jones, E., Oliphant, E., Peterson, P., et al. 2001, http://www.scipy.org/ [Online; accessed 2017-06-01]
Kasuga, T., Balam, D.D., Wiegert, P.A. 2011, EPSC-DPS Joint Meeting 2011, p.69
Kelley, M.S., Lindler, D.J., Bodewits, D., et al., 2013, Icar, 222, 634
Knight, M.M., CBET 4251
Knight, M.M., & Schleicher, D.G. 2016, DPS 2016, 217.02
Kramer, T., Noack, M. 2016, ApJL, 823, L11
Krist, J.E., Hook, R.N., & Stoehr, F. 2011, Proceedings of SPIE, 8127, 81270J
Lamy, P.L., Toth, I., A'Hearn, M.F., et al. 1999, Icar, 140, 424

Lamy, P.L., Toth, I., Fernández, Y.R., et al. 2004, In: Comets II, pp. 223
Lamy, P.L., Toth, I., Weaver, H.A., et al. 2011, MNRAS, 412, 1573
Landolt, A.U. 2009, AJ, 137, 4186
Li, J.-Y., Kelley, M.S.P., Knight, M.M., et al. 2013a, ApJL, 779, L3
Li, J.-Y., Besse, S., A'Hearn, M.F., et al. 2013b, Icar, 222, 559
Li, J.-Y., Samarasinha, N.H., Kelley, M.S.P., et al. 2014, ApJL, 797, L8
Li, J.-Y., Samarasinha, N.H., Kelley, M.S.P., et al. 2016, ApJL, 817, L23
Lomb, N.R. 1976, Ap&SS, 39, 447
Marsden B.G., Sekanina, Z., Yeomans, D.K. 1973, AJ, 78, 211
McKay, A.J., Kelley, M.S.P., Bodewits, D., et al. 2017, ACM, 3b4
Meech, K.J., Hainaut, O.R., & Marsden, B.G., 2004, Icar, 170, 463
Mortier, A., Faria, J.P., Correia, C.M., et al. 2015, A&A, 573, A101 (6pp)
Naidu, S.P., Benner, L.A.M., Brozovic, M., et al. 2016, AAS DPS meeting #48, id.219.05
Neishtadt, A.I., Scheers, D.J., Sidorenko, V.V., et al. 2002, Icar, 157, 205
Pérez, F., Granger, B.E. 2007, Computing in Science & Engineering, 9, 21
Protopapa, S., Sunshine, J.M., Feaga, L.M., et al. 2014, Icar, 238, 191
Rayman M.D., 2003, Space Technol., 23, 185
Rotundi, A., Sierks, H., Della Corte, V., et al. 2015, Science, 347, aaa3905-1
Tricarico, P., & Samarasinha, N. 2011. LPI, 42, 2721
Samarasinha, N.H., & A'Hearn, M.F. 1991, Icar, 93, 194
Samarasinha, N.H., Belton, M.J.S. 1995, Icar, 116, 340
Samarasinha, N.H., Mueller, B.E.A. 2013, ApJL, 775, L10 (6pp)
Samarasinha, N.H., Larson, S.M. 2014, Icar, 239, 168
Scargle, J.D. 1982, ApJ, 263, 835
Sekanina, Z., Brownlee, D.E., Economou, T.E., et al. 2004, Science, 304, 1769
Shelly, F., Brown-Manguso, L., Blythe, M., et al. 2000, IAU Circular, 7396
Smith, J.A., Tucker, D.L., Kent, S., et al. 2002, AJ, 123, 2121
Snodgrass, C., Fitzsimmons, A., Lowry, S.C., et al. 2011, MNRAS, 414, 458
Soderblom, L.A., Boice, D.C., Britt, D.T., et al. 2004, Icar, 167, 4
Sosa, A., Fernández, J.A. 2015, IAU General Assembly, Meeting #29, id.2255583
Southwarth, R.B., & Hawkins, G.S. 1963, Smithsonian Contributions to Astrophysics, 7, 261
Tholen D.J. & Chesley S.R., 2004, BAAS, 36, 1151
Thomas, P.C., Veverka, J., Belton, M.J.S., et al. 2007, Icar, 187, 4
Thomas, P.C., A'Hearn, M., Belton, M.J.S., et al. 2013, Icar, 222, 453
Toth, I., 1999, Icar, 141, 420
van der Walt, S., Colbert, S.C., Varoquaux, G. 2011, ArXiv e-prints, arXiv:1102.1523
Ye, Q.-Z., Brown, P.G., Wiegert, P.A. 2016a, ApJL, 818, L29 (5pp)
Ye, Q.-Z., Hui, M.-T., Brown, P.G., et al. 2016b, Icar, 264, 48
Yeomans D.K. & Chodas P.W. 1989, AJ, 98, 1083
Zechmeister, M. & Kürster, M. 2009, A&A, 496, 577